\documentclass[aps,prl,twocolumn]{revtex4-2}

\usepackage{ulem}

\usepackage{amsmath}
\usepackage{graphicx}
\usepackage[colorlinks=true, linkcolor=blue, citecolor=blue, urlcolor=blue]{hyperref}
\usepackage{bm}
\usepackage{physics}

\usepackage{xcolor}

\def\EK{}
\newcommand{\bmhat}[1]{\hat{\bm{#1}}}

\begin{document}

\title{Correlated emission of electron-current waves}

\author{Shane P. Kelly}
\affiliation{Department of Physics and Astronomy and Mani L. Bhaumik Institute for Theoretical Physics, University of California at Los Angeles, CA 90095}

\author{Eric Kleinherbers}
\affiliation{Department of Physics and Astronomy and Mani L. Bhaumik Institute for Theoretical Physics, University of California at Los Angeles, CA 90095}

\author{Yanyan Zhu}
\affiliation{Department of Physics and Astronomy and Mani L. Bhaumik Institute for Theoretical Physics, University of California at Los Angeles, CA 90095}

\author{Yaroslav Tserkovnyak}
\affiliation{Department of Physics and Astronomy and Mani L. Bhaumik Institute for Theoretical Physics, University of California at Los Angeles, CA 90095}

\begin{abstract}
    Correlated emission of light offers a potential avenue for entanglement generation between atomic spins, with potential application for sensing and quantum memory.
    In this work, we investigate the conditions for the correlated emission by color centers into an electronic bath of conduction electrons.
    Unlike emission into bosonic modes, electrons can absorb energy via two-particle processes across a large range of length scales.
    We find that two length scales are particularly relevant in a clean system: one set by the Fermi velocity and the frequency of the color centers $v_F/\Delta$, and the other set by the Fermi wavelength $\lambda_F \ll v_F/\Delta$.
    Subradiance requires emitters to be spaced at a distance closer than the Fermi wavelength, while superradiance requires spacing less than $\sqrt{\lambda_F v_F/\Delta}$, so long as the emitters are initialized with coherence.
    We show that the emitted current burst has a spiral form, and we discuss the experimental possibility to observe correlated dissipation by color-center qubits coupled to electronic environments.
\end{abstract}

\maketitle

Correlated emission occurs when a collection of localized quantum systems emit energy coherently into an environment.
First identified by R. H. Dicke~\cite{PhysRev.93.99}, the phenomenon has been observed in platforms ranging from cold atomic gases~\cite{PhysRevLett.30.309,PhysRevLett.43.343,doi:10.1126/sciadv.1601231}, quantum dots~\cite{scheibnerSuperradianceQuantumDots2007}, photonic waveguides~\cite{PhysRevX.14.011020,solanoSuperradianceRevealsInfiniterange2017} and nitrogen-vacancy centers~(NVs) in diamond~\cite{bradacRoomtemperatureSpontaneousSuperradiance2017a,angererSuperradiantEmissionColour2018a, wuSuperradiantMaserNitrogenvacancy2021}.
It is important for quantum technologies, both as a potential hazard for error correction~\cite{PhysRevA.96.062337}, and as a resource for entanglement generation~\cite{zouBellstateGenerationSpin2022} and quantum memories~\cite{PhysRevX.7.031024}.

\begin{figure}[!ht]
    \centering
    \includegraphics[width=.94\columnwidth]{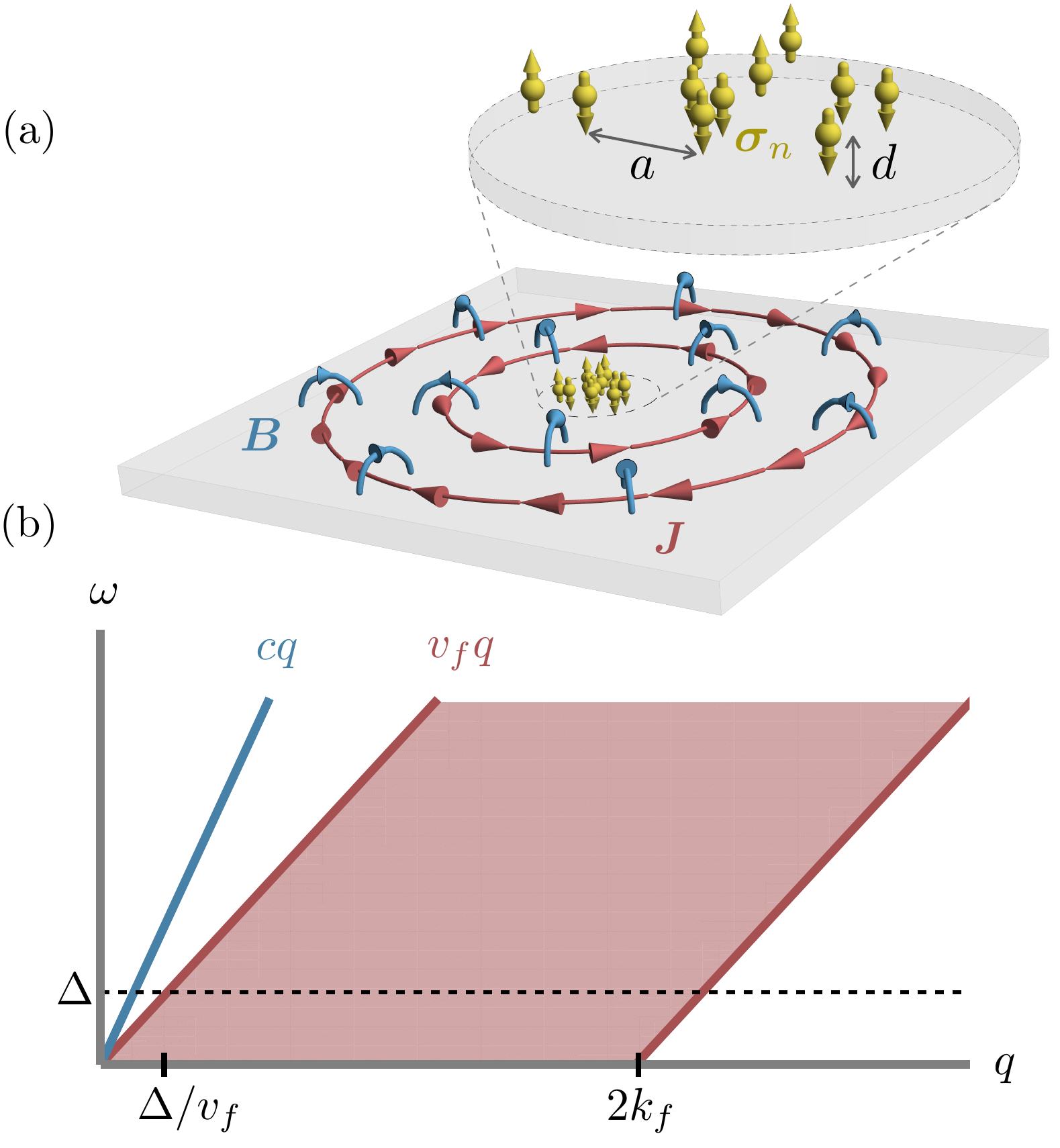}
    \caption{{\EK (a) Sketch of correlated emission of electron-current waves (red) by an ensemble of color centers (yellow) into a bath of conduction electrons (gray).} The Oersted field (blue) couples the electrons to the magnetic dipoles of the color centers. In the magnified view, we show the mean spacing  $a$ between the dipoles $\bm{\sigma}_n$, with $n=1,\ldots, N$, and their height $d$ above the conductor. (b) Sketch of magnetic noise spectrum of the electronic environment (red) which mediates the correlated emission of electron-current waves.  It is characterized by the particle-hole continuum with noise having spectral weight in the shaded region. Qubits couple to a single frequency $\Delta$ of the magnetic noise. For comparison, photonic noise (blue) is peaked on the curve $\omega = c q$ set by the dispersion of light.}
    \label{fig:cartoon}
\end{figure}

Numerous investigations have considered a collection of qubits coupled to photonic environments~\cite{RevModPhys.90.031002}, and recently theoretical work has considered the coupling of qubits to a magnonic environment~\cite{li2024solidstateplatformcooperativequantum}. 
While independent emission is generic, correlated emission can occur if the emitters are spaced closer than the wavelength of the emitted bosonic excitation.
When that is the case, two types of correlated emission are generally identified: subradiance, when the rate of emission is suppressed due to cooperative effects, and superradiance, when the rate of emission is cooperatively enhanced.

In this Letter, we instead study correlated emission into a \textit{fermionic} environment.
Recent experiments~\cite{kolkowitzProbingJohnsonNoise2015,Ariyaratne:2018aa} have observed the dissipative coupling of a single color center, such as the NV center, to the magnetic noise produced by conduction electrons in agreement with theoretical predictions~\cite{langsjoenPRA12}.
In these experiments, single qubits are used as magnetic-field sensors~\cite{taylorHighsensitivityDiamondMagnetometer2008,lee-wongNanoscaleDetectionMagnon2020,doi:10.1126/sciadv.1602429,liNanoscaleMagneticDomains2023} of a condensed-matter environment.
When multiple qubits are available, the condensed-matter environment has the possibility to mediate cooperative effects through dissipation, such as correlated emission, see Fig.~\ref{fig:cartoon}(a). 

A similar cooperative, yet coherent effect, occurs when conduction electrons mediate long-range exchange interactions~\cite{PhysRev.96.99,10.1143/PTP.16.45,PhysRev.106.893} between localized spins.
These coherent interactions may result in spin-glass physics~\cite{RevModPhys.58.801}, characterized by disordered ground states and slow relaxation.
In comparison, color centers are placed outside the electron gas and can be optically controlled and read out, allowing access to far-from-equilibrium quantum states~\cite{PhysRevLett.93.130501}, microscopic probes of spin dynamics, and the tunable weak-coupling regimes relevant for correlated emission~\cite{PhysRev.93.99,zouBellstateGenerationSpin2022,RevModPhys.90.031002}.

Thus, we theoretically consider the possibility of collective emission into an electronic environment.
The primary challenge is that, unlike a bosonic environment such as the electromagnetic vacuum, noise from an electronic gas occurs within a large range of length scales for a fixed emitter frequency.
Electronic noise from conduction electrons is generically captured by the two-particle spectrum of a Fermi gas~(see Fig.~\ref{fig:cartoon}) and is characterized by two length scales: the Fermi wavelength $\lambda_F$ and a length $v_F/\Delta$ depending on the frequency of the emitter $\Delta$ and the Fermi velocity $v_F$.

The main result of this work is the emitter spacing at which subradiance and superradiance will occur due to a two-dimensional electron gas:
Subradiance requires emitters to be spaced closer than the Fermi wavelength, while the requirement for superradiance is generically looser.
In particular, it depends on whether the emitters are initialized in a state with or without coherence between the two levels of the emitter.
If the emitters are initialized without coherence, superradiance requires a spacing on a scale similar to the Fermi wavelength.
While if the emitters are initialized with coherence, superradiance can occur with a spacing $a$ significantly further than the Fermi wavelength so long as $a<\sqrt{\lambda_F v_F/\Delta}$.

\textit{Bosonic vs~fermionic environments.---}
In general, collective emission can occur when an ensemble of localized quantum systems couple to a common $d$-dimensional environment of volume $V$.
As a concrete example, we will focus on spin-half magnetic dipoles, leaving the straightforward generalization to arbitrary two-level systems implicit.
In their ground state, the magnetic dipoles point in the $-\bmhat{z}$ direction.
The magnetic dipoles are described by the Pauli spin operators $\bm{\sigma}_n=(\sigma^{x}_n,\sigma^{y}_n,\sigma^z_n)$, and we model their evolution using the Hamiltonian
\begin{eqnarray}
    H=\frac{\hbar \Delta}{2}\sum_{n=1}^N\sigma^z_n+\frac{\hbar \gamma_e}{2}\sum_{n=1}^N\bm{B}(\bm{r}_n)\cdot\bm{\sigma}_n+H_{\text{E}},
    \label{eq:model}
\end{eqnarray}
where $\hbar\Delta$ is the energy spacing between the two levels of the magnetic dipole, $\bm{B}(\bm{r}_n)$ is the environment-induced magnetic-field operator at the position of the $n$th dipole with $n=1,\ldots,N$, $H_{\text{E}}$ is the Hamiltonian of the environment creating the magnetic fields, and $-\gamma_e$ is the gyromagnetic ratio of the magnetic dipole.
Below, we assume the environment is translationally invariant. {\EK Furthermore, we assume that a direct magnetic dipole-dipole interaction can be neglected, requiring that environment-induced interactions dominate for large enough separations of magnetic dipoles~\footnote{We are effectively requiring that separations $\left|\bm{r}_n-\bm{r}_m\right|$ between dipoles fulfill $\left|\bm{r}_n-\bm{r}_m\right|\gtrsim\sqrt[3]{\hbar \gamma_e^2/\abs{\gamma_{nm}}}$, where the nonlocal decay rate $\gamma_{nm}$ is defined below Eq.~\eqref{eq:dyneqom}.}}  

Under a Markovian weak-coupling approximation, the dynamics of magnetic dipoles are described by a Lindblad master equation containing both coherent mediated interactions and dissipative processes such as excitation, dephasing, and decay.
Here, we consider only decay processes. A more complete treatment is discussed in the Supplemental Material~\cite{SM}. The resulting Lindblad equation in the rotating frame of the dipoles is
\begin{eqnarray}~\label{eq:dyneqom}
    \partial_t \rho = \sum_{nm}\gamma_{nm}\left(\sigma_m^-\rho\sigma_n^+-\frac{1}{2}\{\sigma^+_n\sigma^-_m,\rho\}\right),
\end{eqnarray}
where $\sigma^\pm_n=(\sigma^x_n\pm i\sigma^y_n)/2$ are the Pauli raising and lowering operators, and the nonlocal decay rates are determined by the magnetic-field noise $\gamma_{nm}= \gamma(\bm{r}_n-\bm{r}_m)=\frac{\gamma_e^2}{4}C^{-+}(\Delta,\bm{r}_n-\bm{r}_m)$ at resonant frequency $\Delta$.  The magnetic-field noise at arbitrary frequency $\omega$ is described by the correlation function
\begin{align*}
     C^{-+}(\omega,\bm{r}_n-\bm{r}_m) =\int_{-\infty}^{\infty}dt e^{i\omega t}\left<B^{-}(t,\bm{r}_n)B^{+}(0,\bm{r}_m)\right>
\end{align*}
between the two circular polarizations of the magnetic field $B^{\pm}=B^x\pm iB^y$. 

As a reference, we first consider a bosonic environment described by a collection of bosonic modes $b_{\bm{q}}$ with dispersion $\omega_{\bm{q}}$ and governed by the Hamiltonian $H_{\text{E}}=\sum_{\bm{q}}\hbar\omega_{\bm{q}}b^\dagger_{\bm{q}}b_{\bm{q}}$.
The magnetic fields are linearly related to the bosonic creation and annihilation operators by
\begin{align}
    \bm{B}(\bm{r})= \sum_{\bm{q}} \frac{e^{i\bm{q}\cdot{\bm{r}}}}{\sqrt{V}} \bm{g}_{\bm{q}} b_{\bm{q}} +H.c.,
\end{align}
 such that the magnetic dipoles decay via emission of a boson.
Bosons can only be created at a momentum $\bm{q}$ determined by their energy $\hbar\omega=\hbar\omega_{\bm{q}}$.
Accordingly, the magnetic-field noise is constrained as $C^{-+}(\omega,\bm{q})\propto \delta(\omega_{\bm{q}}-\omega)$, see Fig.~\ref{fig:cartoon}, where $C^{-+}(\omega,\bm{q})=\int d^d\bm{r}e^{-i\bm{q}\cdot\bm{r}}C^{-+}(\omega,\bm{r})$. 

In contrast, an environment of fermionic particles necessarily couples via particle scattering:
the magnetic-field operator is a bosonic operator and must be expanded as a product of an even number of fermion operators.
We consider a noninteracting electron gas described by the Hamiltonian $H_{\text{E}}=\sum_{\bm{k}}\hbar\omega_{\bm{k}}f^\dagger_{\bm{k}}f_{\bm{k}}$, where $f_{\bm{k}}$ is the fermion annihilation operator at wave vector $\bm{k}$ and $\omega_{\bm{k}}=\hbar\bm{k}^2/2m$ is now the electron dispersion with effective electron mass $m$.
In a magnetostatic approximation, the magnetic fields are linearly related to the electron spin and current densities. 
Thus, the magnetic-field operators have, in a homogeneous environment, the form
\begin{eqnarray}
    \bm{B}(\bm{r})=\sum_{\bm{k},\bm{q}}\frac{e^{i\bm{q}\cdot\bm{r}} }{V}\bm{V}_{\bm{k},\bm{k}+\bm{q}} f^{\dagger}_{\bm{k}}f_{\bm{k}+\bm{q}},
\end{eqnarray}
where the matrix elements $\bm{V}_{\bm{k},\bm{k}+\bm{q}}$ are determined by the geometry of the magnetostatic problem, and we have omitted the spin indices. 
Physically, the magnetic-dipole coupling $\propto \bm{\sigma}_n\cdot\bm{B}$ describes processes in which energy is exchanged between the $n$th magnetic dipole and a bosonic electron-hole pair with momenta $\bm{k}+\bm{q}$ and $\bm{k}$.

The kinetics of this process determine the magnetic-field noise governing correlated emission.
The magnetic-field noise at thermal equilibrium takes the form 
\begin{align*}
        C^{-+}\left(\omega, \bm{q}\right)= \int \frac{d^d\bm{k}}{(2\pi)^{d-1}}&V^{-}_{\bm{k},\bm{k}+\bm{q}}V^{+}_{\bm{k}+\bm{q},\bm{k}} n_{\bm{k}}(1-n_{\bm{k}+\bm{q}})  \\ \nonumber
        &\,\,\,\times \delta\left( \omega+\omega_{\bm{k}}-\omega_{\bm{k}+\bm{q}} \right),
\end{align*}
determined by the matrix elements $V^{\pm}_{\bm{k},\bm{k}+\bm{q}}=V^{x}_{\bm{k},\bm{k}+\bm{q}}\pm i V^{y}_{\bm{k},\bm{k}+\bm{q}}$, the Fermi-Dirac occupation function $n_{\bm{k}}$, and the kinetic constraint $\delta\left( \omega+\omega_{\bm{k}}-\omega_{\bm{k}+\bm{q}} \right)$ which restricts the particle at wave vector $\bm{k}$ to absorb only momentum $\hbar\bm{q}$ and energy $\hbar\omega$.
At low temperature and small energy transfer $\hbar\omega \ll E_F$, the magnetic-field noise arises from electron dynamics near the Fermi surface.
The frequencies and wave vectors at which noise can occur are shown in Fig.~\ref{fig:cartoon}.
The largest possible wave vector $\bm{q}$, for small $\hbar \omega$, has magnitude $2k_F$ and is due to an electron scattering between antipodal points on the Fermi surface~\footnote{For an anisotropic Fermi surface, the largest wavenumber $q$ depends on the orientation of $\bm{q}$. Correlated decay sensitive to $k_F$ might then be able to probe the shape of the Fermi surface.}. 
The smallest wave vector has magnitude $\omega/v_F$. 

We now consider a random distribution of dipoles with mean spacing $a$, at a fixed height $d$ above a two-dimensional conductor, see Fig.~\ref{fig:cartoon}(a).
This configuration is characteristic of the experiments in Refs.~\cite{Doherty2013,Ariyaratne:2018aa,kolkowitzProbingJohnsonNoise2015} probing the dynamics of a single NV~\footnote{The NV qubit comprises two levels of an electronic triplet, and the coupling to the magnetic field is generally anisotropic.} at a fixed height above a conductor.
In this setting, the NV predominantly couples to transverse current waves~\cite{PhysRevB.106.L081122,chatterjeeSinglespinQubitMagnetic2022, kelly2024superconductivityenhancedmagneticfieldnoise}: longitudinal waves occupy a negligible region of phase space and are suppressed by the Coulomb interaction.
The coupling-matrix elements between the transverse currents and NVs are determined in the Supplemental Material~\cite{SM} and are 
\begin{eqnarray}
    \bm{V}_{\bm{k},\bm{k}+\bm{q}}=\frac{2\pi e\hbar}{cm}e^{-qd}(\bmhat{q}_{\perp}\cdot\bm{k})(\bmhat{q}+i\bmhat{d}),
\end{eqnarray}
where $\bm{q}$ points in plane; $\bmhat{d}$ points perpendicular to the plane; and $\bmhat{q}_{\perp}=\bmhat{q}\times\bmhat{d}$ is the in-plane unit vector orthogonal to $\bm{q}$. The constants $c$, $m$, and $e$ are the speed of light, effective electron mass, and magnitude of the electron charge.
Finally, we consider the situation in which the NV axes are aligned perpendicular to the plane $\bmhat{z}=\bmhat{d}$.

\begin{figure}[t]
    \centering
    \includegraphics[width=\columnwidth]{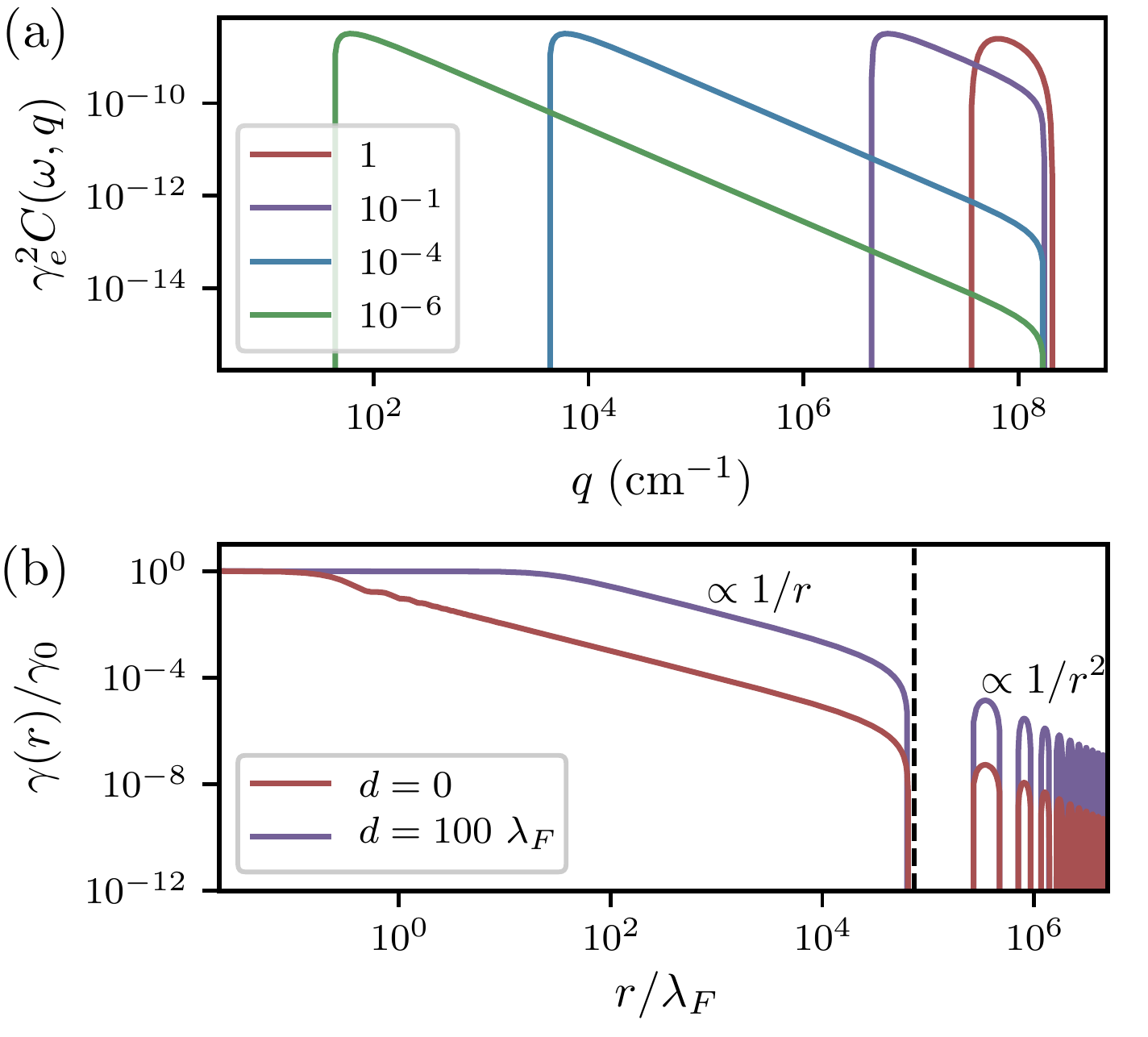}
    \caption{{\EK (a) Magnetic noise spectrum $\gamma_e^2C^{-+}(\omega,q)$ in units of $\text{Hz~cm}^2$ as a function of wave number $q$ for fixed frequency $\omega$.} The legend shows $\hbar\omega/E_F$. The curves were evaluated using $\gamma_e/2\pi=2.8~\text{MHz}/\text{G}$ for the gyromagnetic ratio~\cite{Doherty2013} and $d=0$ for the height. {\EK (b) Nonlocal decay rate $\gamma(r)$  as a function of  magnetic dipole separation $r$.} For the red  line, the height is $d= 0$, while for the purple line $d= 100\,\lambda_F$. 
    The black dashed line indicates the scale $v_F/\Delta$ at which the curves cross over from $1/r$ to an oscillatory $1/r^2$, where we use $\Delta/2\pi=2.87~\text{GHz}$.  In both figures, we use the free electron mass $m=9.1\times 10^{-28}~\text{g}$ and a Fermi velocity $v_F=10^8~\text{cm}/\text{s}$ typical for an electron gas. Accordingly, the Fermi surface is defined by $\lambda_F\approx 10^{-8}~\text{cm}$ and $E_F/\hbar\approx 10^{16}~\text{Hz}$.}
    \label{fig:noise_n_rates}
\end{figure}

The Oersted magnetic-field noise produced by the transverse-current fluctuations~\cite{SM} is shown in Fig.~\ref{fig:noise_n_rates}(a) for zero temperature. 
This noise has spectral weight over a range of wave vectors  $q \in (\omega / v_F, 2 k_F)$.
In this range, the transverse current fluctuations decay as $1/q$.
This Oersted magnetic noise at resonant frequency $\Delta$ induces the decay-rate matrix $\gamma_{nm}$, which we assume to be dominant compared to any intrinsic spin-lattice relaxation of the NVs~\cite{kolkowitzProbingJohnsonNoise2015}.
The decay rate $\gamma_{nm}= \gamma(\bm{r}_n-\bm{r}_m)$ is shown in Fig.~\ref{fig:noise_n_rates}(b) as a function of the dipole separation $r=\left|\bm{r}_n-\bm{r}_m\right|$, where we $\gamma_0=\gamma_{nn}$ is the local decay rate. The characteristic length scales of this function are the dipole height $d$, the Fermi wavelength $\lambda_F$, and $v_F/\Delta$. The local rate scales as $\gamma_0\sim \Delta \gamma_e^2 e^2/\text{max}(\lambda_F,d)\lambda_F c^2$~\cite{SM}.
First, consider the case when the dipoles are at small height $d<\lambda_F$.
In this limit, the decay-rate matrix remains nearly constant for dipole separations $r<\lambda_F$.
At around $r\approx \lambda_F$, we observe wiggles with periodicity $\lambda_F$.  
For larger separations $\lambda_F<r<v_F/\Delta$, the decay-rate matrix decays as $1/r$, before crossing over to $1/r^2$ for $r>v_F/\Delta$ with oscillations of periodicity $2\pi v_F/\Delta$.
When the dipoles are placed at a larger height $d>\lambda_F$, the decay-rate matrix stays constant over a larger separation $r<d$, before crossing over to a $1/r$ decay for $d<r<v_F/\Delta$~\footnote{For finite temperatures, we expect that the length scale $v_F/\Delta$ has to be replaced by $v_F/\max(\Delta,k_BT/\hbar)$.}.

\textit{Subradiance.---}
Subradiance is the phenomenon in which the decay of the collective ensemble is slower than for independent emitters.  If the ensemble is prepared in a state with a single excitation, subradiance occurs when the smallest eigenvalue $\gamma_\text{min}$ of the decay-rate matrix $\gamma_{nm}$ fulfills $\gamma_\text{min}\ll \gamma_0$. 
For two dipoles, $\gamma_\text{min}=\gamma_0-\left|\gamma_{12}\right|\geq 0$, where $\gamma_{12}=\gamma(\left|\bm{r}_1-\bm{r}_2\right|)$ is the non-local rate. 
As the separation $a=\left|\bm{r}_1-\bm{r}_2\right|$ decreases, the non-local rate approaches the local rate, $\gamma_{12}\rightarrow \gamma_0$, and the singlet state becomes dark~(does not decay).
For noise produced by transverse-current fluctuations, this occurs when $a\lesssim \lambda_F$, as shown in Fig.~\ref{fig:noise_n_rates}(b) for $d=0$.
For dipoles placed further from the material $d>\lambda_F$, subradiance occurs for larger spacing $a\lesssim d$. 

These conclusions for two dipoles generalize to a homogeneous collection of dipoles except for one caveat:
For a homogeneous distribution of dipoles with mean spacing $a>\lambda_F$, there is a finite probability that two dipoles have a spacing $r<\lambda_F$.
Thus, even at low densities, there exist subradiant states.
Nonetheless, these subradiant states are statistically rare and localized at the two dipoles with atypical spacing $r\ll a$.
Only when the densities are sufficiently large, $n>\min(\lambda_F^{-2},d^{-2})$, do subradiant states become typical and relevant for collective relaxation dynamics.
This is confirmed in the SM~\cite{SM} via numerical diagonalization of the decay-rate matrix.

\textit{Superradiance.---}
In contrast to subradiance, superradiance can occur when the dipoles are spaced at a distance much larger than either the Fermi wavelength or the height $d$.
In particular, by studying the time dependence of the collective spin $\bm{S}=\frac{\hbar}{2}\sum_n \boldsymbol{\sigma}_n$, we will show that preparing the spin ensemble in an initial state with $\ev{\bm{S}}=\hbar N \hat{\boldsymbol{x}}/2$ rather than $\ev{\bm{S}}=\hbar N \hat{\boldsymbol{z}}/2$ will lead to a relaxed requirement for superradiant enhancement. While exact simulation of the dissipative quantum dynamics of Eq.~\eqref{eq:dyneqom} is computationally prohibitive, properties of early time dynamics are easy to compute and have been used to determine conditions required for superradiance~\cite{PhysRevLett.125.263601,Masson:2022aa}.

First, consider the typical situation when all dipoles are prepared in their excited state, such that $\ev{\bm{S}}=\hbar N \hat{\boldsymbol{z}}/2$.
The initial total decay rate $R=-\partial_t \ev{S^z}/\hbar\vert_{t=0}$ is always equal to the total decay rate for independent emission $R=N\gamma_{0}$ regardless of whether collective effects are present or not.
Collective emission, at early times, is instead distinguished by the initial change in the total decay rate $\dot{R}=-\partial^2_t \ev{S^z}/\hbar\vert_{t=0}$.
For independent emission, the decay process slows and $\dot{R}<0$, while for superradiance, the decay of the first excitation accelerates the decay of the second such that $\dot{R}>0$.
In the case that the dipoles are closely packed and $\gamma_{nm}=\gamma_0$, the change in the total decay rate $\dot{R}$ is positive and scales with $N^2$.

When the dipoles are placed in random locations spanning an area $A$ and with a density $n$, the superradiant enhancement is given by~\cite{SM}
\begin{eqnarray}
    \frac{\dot{R}}{N\gamma_{0}^2} = \pi \lambda_{\text{SR}}^2 n-1,\label{eq:6}
\end{eqnarray}
where we assumed a homogeneous distribution of dipoles and $\lambda_{\text{SR}}$ sets the superradiance length scale at which correlated emission occurs:
\begin{eqnarray}
    \pi\lambda_{\text{SR}}^2=\frac{1}{C^{-+}(\Delta)^2}\int \frac{q dq}{2\pi} C^{-+}(\Delta,q)^2.
\end{eqnarray}
Here, $C^{-+}(\Delta) =(2\pi)^{-1} \int q dq C^{-+}(\Delta,q)$ is the local spectral density at a frequency $\Delta$, and we have assumed the noise is isotropic in $\bm{q}$. 

\begin{figure*}[ht!]
    \centering
    \includegraphics[width=.8\textwidth ]{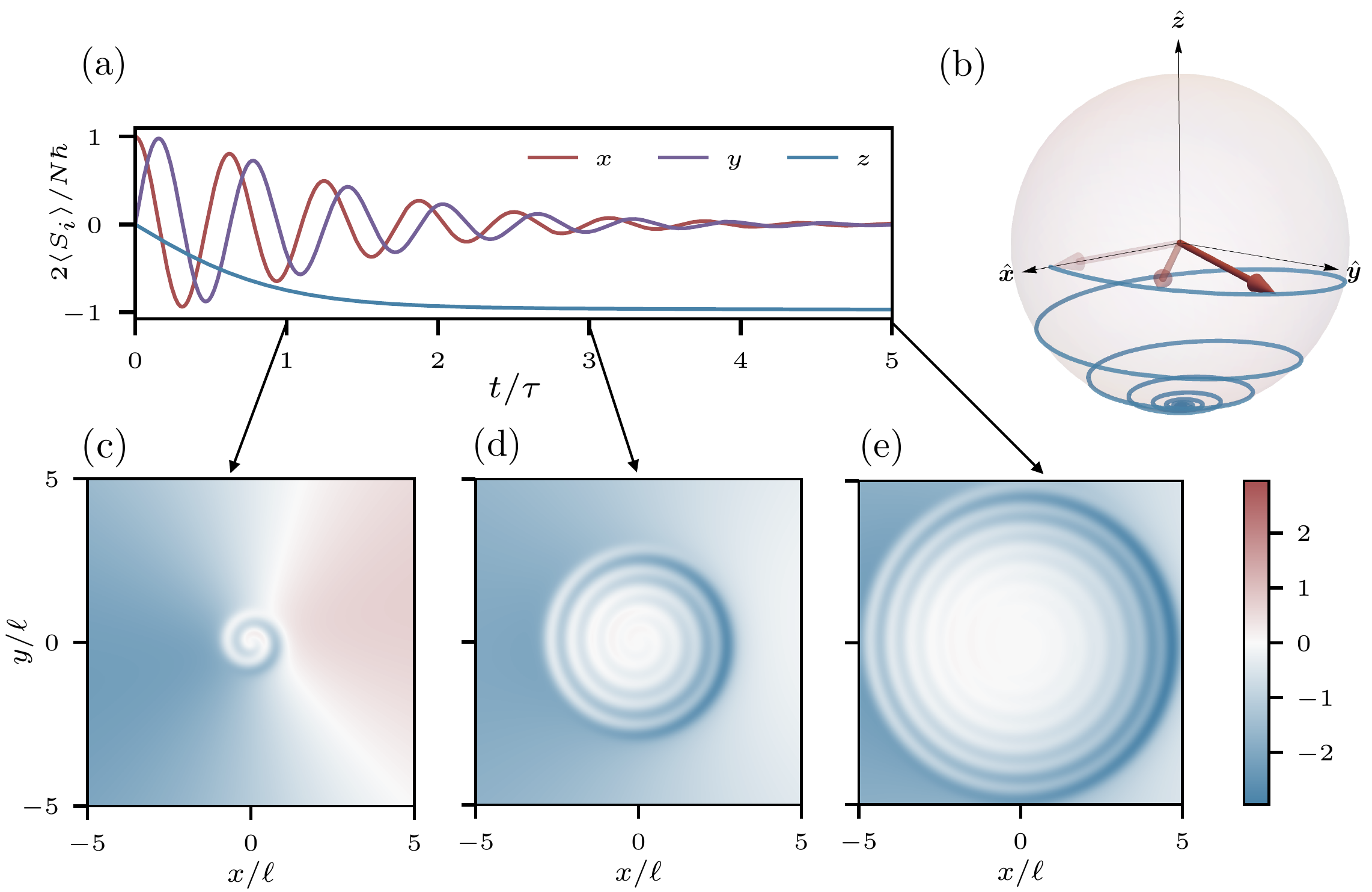}
    \caption{{\EK The decay of the macrospin, initialized with $\ev{\bm{S}}=\hbar N\hat{\boldsymbol{x}}/2$, as a function of time for the components $\ev{S_i}$ with $i\in{\{x,y,z\}}$ in (a), and the corresponding trajectory on the Bloch sphere in (b). (c)-(e) Spatial distribution  of emitted electron current waves at distinct times $t\in \{\tau,3\tau,5\tau\}$ with $\tau=2/N\gamma_0$. The $x$ and $y$ axes have units of $\ell =v_F\tau$. The color bar indicates the azimuthal component of the far-field current distribution $\rho^2{J}_\varphi(\bm{\rho},t)$ in units of $Ne^2v_F^2 \gamma_e m/4\pi \hbar c$, where clockwise (counterclockwise) current flow is colored blue (red).} 
    To clearly distinguish oscillatory and decaying features, the figure assumes extreme parameters, where the macrospin comprised of $N=20$ dipoles is precessing with frequency $\Delta=100\gamma_0$, placed at a height $d=10^{-3}v_F/\gamma_0$ away from a conductor with Fermi energy $E_F=10^6 \hbar \gamma_0/2$. A video showing the evolution can be found at \cite{SM}.
    }
    \label{fig:emitted_currents}
\end{figure*}
The magnetic-field noise produced by the transverse current fluctuations allows for the approximation $C^{-+}(\Delta,q)\propto 1/q$ for $q\in(\Delta/v_F,2k_F)$ and $0$ elsewhere, see Fig~\ref{fig:noise_n_rates}, where we consider the limit $d=0$ when NVs are close to the material.
In this case, the superradiance length is given by $\lambda_{\text{SR}}^2\sim \lambda_F^2 \ln(4E_F/\hbar\Delta)$, where we have assumed $\lambda_F \ll v_F/\Delta$.
Thus, a superradiant acceleration, $\dot{R}>0$, requires that the density of dipoles is larger than $n>\lambda_{\text{SR}}^{-2}/\pi\sim\lambda_F^{-2}/\ln(4E_F/\hbar\Delta)\pi$.
In comparison to subradiance, which requires dipole densities $n\gtrsim \lambda_F^{-2}$, superradiant acceleration occurs at lower densities, albeit by a logarithmic factor $\ln(4 E_F/\hbar\Delta)$~(about 10 for typical NV frequencies and metallic Fermi surfaces). While the requirement on NV densities $n$ seems unrealistic in metallic systems, it might be more promising in doped semiconductors, where the Fermi wavelength $\lambda_F$ is longer and much more tunable.

A more relaxed requirement for collective effects can be observed when the magnetic dipoles are prepared in a state with coherence between the excited and ground state.
Consider an initial state with $\ev{\bm{S}}=\hbar N \hat{\boldsymbol{x}}/2$, easily initialized by a $\pi/2$ pulse, in which each dipole is in an equal superposition of the excited state and the ground state $\left(\ket{-}+\ket{+}\right)/\sqrt{2}$.
For independent emission, the initial total decay rate scales linearly with the number of emitters $R\propto N$.
While for a constant decay-rate matrix $\gamma_{nm}=\gamma_0$, the initial total decay rate scales quadratically $R\propto N^2$.
In a generic setting, we obtain $R=\frac{1}{2}\sum_k\gamma_{kk}+\frac{1}{4}\sum_{n\neq m}\gamma_{nm}$, and 
when the dipoles are distributed randomly in a region of linear size $\sqrt{A}$, it is given by
\begin{eqnarray}\label{eq:Rxhomo}
    R\approx\frac{N}{2}\gamma_0+\frac{N\pi n}{2}\int_0^{\sqrt{A}} rdr \gamma(r),
\end{eqnarray}
where we have approximated the sum over dipole locations via a disorder average over their distribution~\cite{SM}.

For dipoles coupled to transverse current fluctuations and placed close to the sample $d<\lambda_F$, the decay rate follows $\gamma(r)\propto 1/r$ for $r\in (\lambda_F,v_F/\Delta)$.
Thus, so long as $\lambda_F<\sqrt{A}<v_F/\Delta$, there is a superradiant enhancement of the total decay rate $R\sim N n\gamma_0 \lambda_F \sqrt{A}$ scaling with the size of the region $\sqrt{A}$ occupied by the dipoles.
For $\sqrt{A}>v_F/\Delta$, the integrand in Eq.~\eqref{eq:Rxhomo} becomes highly oscillatory when $r>v_F/\Delta$. Assuming these long-distance oscillations are suppressed, for example by disorder,  we estimate $R\sim N n\gamma_0\lambda_F v_F/\Delta$ for $A\rightarrow \infty$~\cite{SM}. 
Therefore, if the average spacing between the dipoles is smaller than 
\begin{equation}
    \lambda_{SR}'=\sqrt{\lambda_F\, v_F/\Delta},   \label{eq:9}
\end{equation}
we expect a collective enhancement of the decay process scaling with $n \lambda_{\text{SR}}^{'2}$.
Note that for a quadratic dispersion, $\lambda_{\text{SR}}'$ is the de-Broglie wavelength for an electron at energy $\hbar\Delta$: approximately $1~\mu$m for a free electron and $\Delta\approx 1$ GHz.
For dipoles further from the sample $d>\lambda_F$, the spacing must be smaller than $\lambda_{\text{SR}}'\approx \sqrt{d v_F/\Delta}$.

Note that, while the first condition, $\dot{R}>0$, is sufficient for a superradiant burst, it may not be necessary.
It is conceivable that at early times $\dot{ R}<0$ due to a slow development of coherence, but at intermediate times, coherence could become sizable leading to superradiance.
To resolve this possibility, an experiment or a more elaborate many-body method is required. Finally, we remark that these initial-time  criteria of superradiance from Eqs.\eqref{eq:6}-\eqref{eq:9} are unchanged when we include the coherent interactions mediated by the fermionic environment (in the secular approximation) and pure dephasing~\cite{SM}.

\textit{Emitted Currents.---}
In the ballistic electronic environment, energy is carried away by a redistribution of electrons around the Fermi surface.
This redistribution forms transverse-current waves, which we determine by linear response to the magnetic fields produced during the decay of the dipoles. We obtain
\begin{align}
\bm{J}(\bm{\rho},t) = \int {dt'}{\bm{\chi}}(\bm{\rho},t-t')\ev{\bm{S}(t')}, 
\end{align}
where $\bm{J}$ is the 2D current density at position $\bm{\rho}$, and ${\bm{\chi}}$ is the tensorial response function calculated in the Supplemental Material~\cite{SM}. For simplicity, we consider the magnetic dipoles to all be located at the same spot, such that they form a collective macrospin $\ev{\bm{S}}$ 
that undergoes classical Landau-Lifshitz damping, see Fig.~\ref{fig:emitted_currents}(a)-(b).

In Fig.~\ref{fig:emitted_currents}(c)-(e), we show the spatial dependence of the emitted transverse-current waves for three distinct times. The colors indicate the azimuthal component ${J}_\varphi$ of the current, which dominates ${J}_\rho$ for large distances $\rho \gg v_F/\Delta$. 
We observe the  formation of a pattern of Archimedean spirals with two arms. 
These two spiral arms merge at the origin (the position of the macrospin) in a way that ensures  $\grad \cdot \bm{J} =0$. In addition, loop currents are superimposed on the spirals, which favor an overall clockwise current direction in accordance with Lenz's law.
The envelope of the emitted azimuthal current waves moves ballistically outwards with Fermi velocity $v_F$. The width $\propto v_F/ N\gamma_0 $ and radial periodicity $2\pi v_F/\Delta$ are inherited from the decay rate of the macrospin $\propto N \gamma_0$ and the dipole frequency $\Delta$, respectively.

\textit{Discussion.---}
We discuss a few novelties relevant to the superradiance of color centers coupled to current fluctuations. 
First, the height dependence provides an additional control: color centers separated by less than their height can exhibit collective effects due to magnetostatic coupling.
Thus, even a paramagnet, with very short-range magnetic fluctuations, could induce collective relaxation simply due to the nonlocality of the magnetostatic coupling. 
Furthermore, materials often display both disorder and interaction effects, which might modify length scales at which correlated emission can occur. 
For example, the mean free path $l$ in many common materials is shorter than $v_F/\Delta$, and the condition for superradiance will be modified: likely requiring $a<l$. 
In addition, realistic Fermi-liquid effects will likely not qualitatively modify the above predictions, even though a strong enough interaction could theoretically produce a transverse-current zero-sound mode~\cite{silin1958oscillations,gor1963feasibility}. 

For future work, it would be interesting to investigate collective effects in color center ensembles induced by different solid-state environments, where long-range correlations—such as those set by the diffusion length of conserved quantities—can give rise to effective superradiant length scales. Moreover, an interesting open question is whether and how such collective effects can survive strong inhomogeneous broadening~\cite{sunPRL19,zhangPRA18} in the color center ensemble when dynamical decoupling procedures~\cite{deLangeScience10} are employed.

\textit{Acknowledgments-}  We appreciate discussions with Shu Zhang and Jamir Marino.  This work is supported by National Science Foundation under Grant No. DMR-2049979.

\bibliography{refs.bib}

\end{document}


\title{Supplemental material for: Correlated emission of electron-current waves}
\author{Shane P. Kelly}
\author{Eric Kleinherbers}
\author{Yanyan Zhu}
\author{Yaroslav Tserkovnyak}

\affiliation{Department of Physics and Astronomy and Mani L. Bhaumik Institute for Theoretical Physics, University of California, Los Angeles, CA 90095}
\maketitle
\onecolumngrid
\vspace{-3em}

\tableofcontents
\newcommand{\xrm}{x}
\newcommand{\yrm}{y}
\newcommand{\zrm}{z}

\section{Model}
In the Supplemental Material, we will consider a slightly more generic setting in which the ground state of the $n$th magnetic dipole points in a generic direction $\bmhat{\zrm}_n$, and defines a coordinate system $\bmhat{\xrm}_n,\bmhat{\yrm}_n,\bmhat{\zrm}_n$.
The magnetic dipoles are described by Pauli spin operators, $\bm{\sigma}_n=(\sigma^{\xrm}_n,\sigma^{\yrm}_n,\sigma^\zrm_n)$, whose dynamics are described by a Hamiltonian
\begin{eqnarray}\label{eq:model}
    H=\frac{\hbar\Delta}{2}\sum_n\sigma^{\zrm}_n+\gamma_e\frac{\hbar}{2}\sum_n\bm{B}(\bm{r}_n)\cdot\bm{\sigma}_n+H_{\text{E}}
\end{eqnarray}
where $\hbar\Delta$ is the energy spacing of the two levels, $-\gamma_e$ is the gyromagnetic ratio of the magnetic dipole, $\bm{B}(\bm{r}_n)$ is the magnetic-field operator at the position of the $n$th dipole, and $H_{\text{E}}$ is the Hamiltonian of the environment determining the dynamics of the magnetic fields in the absence of the dipoles.

\subsection{Projection of the NV triplet to two levels}
In this section, we note that the magnetic levels of the NV are comprised of a spin-$1$ triplet and experiments are typically performed on two of the three levels.
While the projection to two levels does not yield a spin-half dipole, the result can still be written in terms of Pauli operators.
The spin-$1$ operator $\bm{S}_n$ of the $n$th NV center can be written in the eigenbasis $\{\ket{0}, \ket{\Uparrow}, \ket{\Downarrow}\}_n$.  Projecting onto the two-level subset with $\ket{+}=\ket{\Uparrow}$ and $\ket{-}=\ket{0}$, we can identify the Pauli matrices $\boldsymbol{\sigma}_n$ through the projection $\bm{S}_n \rightarrow \frac{\hbar}{2} \left(\sqrt{2} \sigma_n^\xrm,\sqrt{2}\sigma_n^\yrm,1+\sigma_n^\zrm\right)$. 
Hence, the coupling to the magnetic field
does not generically yield the Hamiltonian $\frac{\hbar\gamma_{\text{NV}}}{2}\bm{\sigma}\cdot\bm{B}$ from Eq.~\eqref{eq:model}.
Instead, we obtain 
\begin{eqnarray}
    \gamma_{\text{NV}} \bm{S}_n\cdot\bm{B}(\bm{r}_n)\rightarrow \frac{\hbar\gamma_{\text{NV}}}{2}\sigma^\zrm_nB^\zrm(\bm{r}_n)+\frac{\hbar\gamma_{\text{NV}}}{\sqrt{2}}\left(\sigma^{\xrm}_nB^{\xrm}(\bm{r}_n) + \sigma^{\yrm}_nB^{\yrm}(\bm{r}_n) \right),
\end{eqnarray}
which has an anisotropy between $\zrm$ and $\xrm,\yrm$. Accordingly, in Eq.~\eqref{eq:decayratesfirst}, the replacement $\gamma_e\rightarrow\sqrt{2}\gamma_{\text{NV}}$ should be made to determine the decay (and excitation) rates,  while the replacement $\gamma_e\rightarrow \gamma_{\text{NV}}$ should be made to determine the dephasing rates.
Furthermore, if we prepare the system in the polarized state $\left(\ket{+}+\ket{-}\right)/\sqrt{2}$, the NV spin is not polarized along $x$, but partially points in the $z$ direction $\ev{\bm{S}_n}=\hbar (1/\sqrt{2},0,1/2)$. Similarly, we obtain  $\ev{\bm{S}_n}=0$ in the ground state $\ket{-}$.

\subsection{Bosonic and Fermionic Environments}
We consider two generic environments of volume $V$, one bosonic and the other fermionic, both with dispersion $\hbar \omega_k$.
Both environments are quadratic with the Hamiltonians
\begin{eqnarray}
    H_b=\sum_{\bm{k}}\hbar \omega_k b^{\dagger}_{\bm{k}}b_{\bm{k}} ,\,\,\, 
    H_f=\sum_{\bm{k}}\hbar \omega_k f^{\dagger}_{\bm{k}}f_{\bm{k}}
\end{eqnarray}
where $b_{\bm{k}}$ and $f_{\bm{k}}$ are the bosonic and fermionic annihilation operators corresponding to plane waves with wave vectors $\bm{k}$.
Spin indices could be included, but are not important for the discussion below.

Bosons and fermions couple differently to the magnetic dipole: 
For bosons, we consider a linear coupling in which the magnetic field is proportional to the boson creation and annihilation operators 
\begin{eqnarray}
    \bm{B}(\bm{r})=\sum_{\bm{k}}\frac{\bm{g}_{\bm{k}}(\bm{r})}{\sqrt{V}} b_{\bm{k}}+h.c.
\end{eqnarray}
Such coupling is generic in quantum optics when the dipoles couple to boson modes.
It is also relevant for NV centers coupled to magnons~\cite{PhysRevB.111.064424,PhysRevLett.121.187204}. 
In contrast, the fermionic environment must couple quadratically as the magnetic field is a bosonic operator:
\begin{eqnarray}
    \bm{B}(\bm{r})=\sum_{\bm{k},\bm{k'}}\frac{\bm{V}_{\bm{k},\bm{k'}}(\bm{r})}{V} f^{\dagger}_{\bm{k}}f_{\bm{k'}},
\end{eqnarray}
where $V$ is the $d$-dimensional volume in which the plane waves are quantized.

\subsubsection{Translation Symmetry}
For a homogeneous coupling, the matrix elements satisfy 
\begin{equation}\label{eq:homobservable}
    \begin{split}
        \bm{g}_{\bm{k}}(\bm{r}+\bm{\delta})&=\bm{g}_{\bm{k}}(\bm{r})e^{i\bm{k}\cdot\bm{\delta}} \\ 
        \bm{V}_{\bm{k},\bm{k'}}(\bm{r}+\bm{\delta})&= \bm{V}_{\bm{k},\bm{k'}}(\bm{r})e^{i\left( \bm{k'}-\bm{k} \right)\cdot\bm{\delta}},
    \end{split}
\end{equation}
such that $\bm{g}_{\bm{k}}(\bm{r})=\bm{g}_{\bm{k}}(0)e^{i\bm{k}\cdot\bm{r}}$ and similarly for the fermion matrix elements. Thus, the magnetic fields at wave vector $\bm{q}$, $\bm{B}(\bm{q})=\int_V d\bm{r} e^{-i\bm{q}\cdot\bm{r}}\bm{B}(\bm{r})$ are
\begin{equation}
    \begin{split}
        \bm{B}(\bm{q}) &= \sqrt{V}\left(\bm{g}_{\bm{q}}b_{\bm{q}}+\bm{g}^{*}_{-\bm{q}}b^{\dagger}_{-\bm{q}}\right)  \\
        \bm{B}(\bm{q}) &= \sum_{\bm{k}}\bm{V}_{\bm{k},\bm{k}+\bm{q}}f^{\dagger}_{\bm{k}}f_{\bm{k}+\bm{q}},
    \end{split}
\end{equation}
where we have defined $\bm{g}_{\bm{q}}=\bm{g}_{\bm{q}}(\bm{r}=0)$ and $\bm{V}_{\bm{k},\bm{k}+\bm{q}}=\bm{V}_{\bm{k},\bm{k}+\bm{q}}(\bm{r}=0)$ for bosons and fermions respectively.
The corresponding real-space magnetic-field operators are given in the main text.

\subsubsection{Rotation Symmetry}
We now assume the environment is also symmetric around an axis $\bmhat{d}$, which is generically not the orientation of the dipoles $\bmhat{\zrm}_n$.
In the case of NVs placed above a conductor, $\bmhat{d}$ is the direction perpendicular to the plane of the material.
If $\bmhat{x}$, $\bmhat{y}$ and $\bmhat{d}$ form a right-handed basis, then the magnetic fields $B^{\pm}(\bm{q})=B^x(\bm{q})\pm iB^y(\bm{q})$ transform as $B^{\pm}(\bm{q})\rightarrow B^{\pm}(\bm{q}_\phi)e^{i\phi}$ under a rotation by $\phi$ around the $\bmhat{d}$ axis~(where $\bm{q}_{\phi}$ is the vector $\bm{q}$ rotated around $\bmhat{d}$ by $\phi$).
Thus, the boson couplings will satisfy
\begin{equation}
    \begin{split}\label{eq:rotsym_boson}
        g^{\pm}_{\bm{q}_\phi}&=g^{\pm}_{\bm{q}}e^{\pm i\phi}\\ 
        g^{d}_{\bm{q}_\phi}&=g^{d}_{\bm{q}},
    \end{split}
\end{equation}
and the fermion couplings will satisfy
\begin{equation}\label{eq:rot_fermion}
    \begin{split}
        V^{\pm}_{\bm{k}_\phi,\bm{k}_\phi+\bm{q}_\phi}&= V^{\pm}_{\bm{k},\bm{k}+\bm{q}}e^{\pm i\phi} \\
        V^{d}_{\bm{k}_\phi,\bm{k}_\phi+\bm{q}_\phi}&= V^{d}_{\bm{k},\bm{k}+\bm{q}}.
    \end{split}
\end{equation}

\subsection{Oersted coupling to 2D current Fluctuations}
For a concrete example, we consider the situation when the magnetic dipoles are located in a two-dimensional plane at a height $d$ above a two-dimensional conductor.
Consistent with previous estimates~\cite{kelly2024superconductivityenhancedmagneticfieldnoise}, we neglect couplings to magnetization noise produced by electron-spin fluctuations and only consider magnetic fields produced by current fluctuations.
The two-dimensional current density has the form
\begin{eqnarray}
    \bm{J}(\bm{q})=\int_V d\bm{\rho}e^{-i\bm{q}\cdot\bm{\rho}}\bm{J}(\bm{\rho})=-\frac{e\hbar}{m}\sum_{\sigma,\bm{k}} \left(\bm{k}-\frac{\bm{q}}{2}\right)f^{\dagger}_{\sigma,\bm{k}-\bm{q}}f_{\sigma,\bm{k}},
    \label{eqn: current operator}
\end{eqnarray}
where $e$ is the magnitude of the electron charge. 
The corresponding magnetic field at wave vector $\bm{q}$ is given by~\cite{kelly2024superconductivityenhancedmagneticfieldnoise}
\begin{eqnarray}\label{eq:magpropcross}
    \bm{B}(\bm{r})=\int_{\mathcal{R}_2}\frac{d^2\bm{q}}{(2\pi)^2}e^{i\bm{q} \cdot \bm{r}}\bm{G}(\bm{q})\cross\bm{J}(\bm{q})
\end{eqnarray}
where
\begin{eqnarray}
    \bm{G}(\bm{q})=\frac{2\pi}{c}e^{-qd}\left(i\bmhat{q}-\text{sign}(d)\bmhat{d}\right),
    \label{eqn: Biot-Savart}
\end{eqnarray}
where $\bmhat{d}$ is the direction perpendicular to the plane, and $\text{sign}(d)$ is determined by the side of the plane being considered.

Defining a right-handed basis by $\bmhat{q}_{\perp},\bmhat{q},\bmhat{d}$, we decompose the current into a longitudinal component $\bmhat{q}\cdot \bm{J}(\bm{q})$ and a transverse component $\bmhat{q}_{\perp}\cdot \bm{J}(\bm{q})$.
The Fermi gas only supports longitudinal current in a small region of phase space near $q\approx \omega/v_F$ and $q\approx 2k_F$.
Furthermore, the Coulomb interaction suppresses longitudinal fluctuations for frequencies below the plasma frequency.
Accordingly, we only consider contributions from transverse fluctuations.
The magnetic field from this component is
\begin{eqnarray}
    \bm{B}(\bm{q})=\frac{2\pi e\hbar}{cm}e^{-qd}(\bmhat{q}+i\bmhat{d})\sum_{\sigma,\bm{k}}\bmhat{q}_{\perp}\cdot\bm{k} f^{\dagger}_{\sigma,\bm{k}}f_{\sigma,\bm{k}+\bm{q}},
\end{eqnarray}
where we have assumed $d>0$ and evaluated the cross product in Eq.~\eqref{eq:magpropcross} using only the transverse component $\left(\bmhat{d}-i\bmhat{q}\right)\cross \bmhat{q}_\perp=(\bmhat{q}+i\bmhat{d})$.
Thus, the coupling matrix is given by
\begin{eqnarray}
    \bm{V}_{\bm{k},\bm{k}+\bm{q}}(0)=(\bmhat{q}+i\bmhat{d})\frac{2\pi e\hbar}{cm}e^{-qd}(\bmhat{q}_{\perp}\cdot\bm{k}),
\end{eqnarray}
and satisfies the rotational symmetry relations in Eq.~\eqref{eq:rot_fermion}.

\section{Environment Fluctuations}
The magnetic-field fluctuations $C^{\alpha\beta}(\omega,\bm{q})$ are qualitatively different for fermions vs bosons.
At a fixed frequency, and in thermal equilibrium, bosonic fluctuations are significantly more constrained:
\begin{eqnarray}
    C^{\alpha\beta}\left(\omega, \bm{q}\right) = 2\pi \left[g^{\alpha}_{\bm{q}}g^{*\beta}_{\bm{q}}(1+n_{\bm{q}})\delta\left( \omega-\omega_{\bm{q}} \right) + g^\beta_{-\bm{q}}g^{*\alpha}_{-\bm{q}}n_{-\bm{q}} \delta\left( \omega+\omega_{-\bm{q}} \right)\right],
\end{eqnarray}
where $n_{\bm{q}}$ is the Bose-Einstein distribution.
In contrast, fermion fluctuations occur for a continuum of wave vectors for a fixed frequency,
\begin{eqnarray}~\label{eq:cqw_fermion_general_sup}
    C^{\alpha\beta}\left(\omega, \bm{q}\right) = \int \frac{d^d\bm{k}}{(2\pi)^{d-1}}V^{\alpha}_{\bm{k},\bm{k}+\bm{q}}V^{\beta}_{\bm{k}+\bm{q},\bm{k}}n_{\bm{k}}(1-n_{\bm{k}+\bm{q}})\delta\left( \omega+\omega_{\bm{k}}-\omega_{\bm{k}+\bm{q}} \right) + F^\alpha F^\beta(2\pi)^{d+1}\delta(\omega)\delta(\bm{q}),
\end{eqnarray}
where $n_{\bm{k}}$ is now the Fermi-Dirac distribution and $F^\gamma=\int d^d\bm{k}V^\gamma_{\bm{k},\bm{k}}n_{\bm{k}}/(2\pi)^d$.
The last term is the static homogeneous response and does not contribute to the dynamics of the dipoles.

For simplicity, we first assume momentum- and direction-independent couplings $V^{\alpha}_{\bm{k},\bm{k}+\bm{q}}=\mathcal{V}$.
In this case, the integral for the fermion fluctuations in Eq.~\eqref{eq:cqw_fermion_general_sup} can be evaluated at zero temperature as
\begin{align}\label{eq:simple}
    C^{\alpha\beta}\left(\omega, \bm{q}\right) &=
    \begin{cases}
        \frac{\sqrt{2}\mathcal{V}^2m^{3/2}}{\pi \hbar^2 q}\left(\sqrt{E_F-E(\omega,q)}-\sqrt{E_F-\hbar\omega-E(\omega,q)}\right)&,\, E(\omega,q)<E_F-\hbar\omega \\
        \frac{\sqrt{2}\mathcal{V}^2m^{3/2}}{\pi \hbar^2 q}\sqrt{E_F-E(\omega,q)}&,\, E_F-\hbar\omega<E(\omega,q)<E_F \\ 
        0&,\, E_F<E(\omega,q)
    \end{cases}
\end{align}
where 
\begin{eqnarray}
    E(\omega,q)= \frac{\hbar^2}{8m}\frac{\left(q_{\omega}^2-q^2\right)^2}{q^2} \text{and } q_{\omega}=\sqrt{\frac{2m\omega}{\hbar}},
\end{eqnarray}
and we have assumed a dispersion $\omega_k = \hbar k^2/2m$.
At small frequency $\hbar\omega\ll E_F$, the fluctuations are finite in an interval determined by $E(\omega,q)<E_F$  or $\omega/v_F<q<2 k_F$.
At zero temperature, $C^{\alpha\beta}\left(\omega, \bm{q}\right)=0$ for $\omega<0$. 

\subsection{Magnetic-field noise from current fluctuations}
The magnetic-field fluctuations produced by the current can also be evaluated exactly for an isotropic 2D Fermi surface:
\begin{eqnarray*}
    C^{\alpha\beta}(\omega,\bm{q})=\left(\bmhat{q}_\alpha+i\bmhat{d}_{\alpha}\right)\left(\bmhat{q}_\beta-i\bmhat{d}_{\beta}\right)e^{-2dq}\left( \frac{2e v_f}{c} \right)^2\frac{\pi m}{\hbar }\frac{k_F}{q}C(\omega,q)
\end{eqnarray*}
where
\begin{equation}
    C(\omega,q)=\frac{2}{3E_F^{3/2}}\left[ \left( E_F-E(\omega,q) \right)^{3/2}-\left( E_F-\hbar\omega-E(\omega,q) \right)^{3/2} \right]
\end{equation}
if $E(\omega,q)<E_F-\hbar\omega$, and has a similar generalization for $E(\omega,q)>E_F$ and $E_F-\hbar\omega<E(\omega,q)<E_F$ as in Eq.~\eqref{eq:simple}.
The main difference is the $3/2$ versus $1/2$ exponent.
Note that we have included a factor of 2 for the spin degrees of freedom.

In the range  $\omega/v_F\ll q\ll k_\text{F}$, we can approximate $E(\omega,\bm{q})\approx \hbar^2q^2/8m$ and make an expansion in both $E(\omega,q)\ll E_F$ and $E(\omega,q)+\hbar\omega \ll E_F$.
This yields 
\begin{align}\label{eq:fluctuations}
    C(\omega,q)\approx\frac{\hbar\omega}{E_F}.
\end{align}

\subsubsection{Role of NV Orientation}
We now determine how the NV orientations affect the relaxation rates at resonant frequency $\omega=\Delta$. As above, we consider the dipoles to point in the direction $\bmhat{\zrm}_n$, with $\bmhat{\xrm}_n,\bmhat{\yrm}_n,\bmhat{\zrm}_n$ forming a right-handed basis. 
The relaxation rates 
\begin{align}~\label{eq:decayratesfirst}
\gamma_{nm}=\frac{\gamma_e^2}{4}\int \frac{d\bm{q}}{(2\pi)^2} e^{i \bm{q}\cdot( \bm{r}_n-\bm{r}_m)}C^{-+}(\Delta,\bm{q})
\end{align}
are then determined by 
\begin{equation}\label{eq:noisepm}
    C^{-+}(\Delta,\bm{q})= F_{nm}e^{-2dq}\left( \frac{2e v_f}{c} \right)^2\frac{\pi m}{\hbar }\frac{k_F}{q}C(\Delta,q),
\end{equation}
where
\begin{equation}
    F_{nm}=\left(\bmhat{\xrm}_n-i\bmhat{\yrm}_n\right)\cdot\left(\bmhat{q}+i\bmhat{d}\right)\left(\bmhat{\xrm}_m+i\bmhat{\yrm}_m\right)\cdot\left(\bmhat{q}-i\bmhat{d}\right)
\end{equation}
contains the orientation dependence. In the simplest configuration, NVs point perpendicular to the plane $\bmhat{\zrm}_n=\bmhat{d}$, and we find $F_{nm}=1$. A numerical evaluation of the magnetic noise is shown in Fig.~2(a) of the main text.

 A more interesting example is when the dipoles point parallel to the plane. Taking $\bmhat{x},\bmhat{y}$ and $\bmhat{d}$ to form a right-handed basis, then the parallel configuration is described by $\bmhat{\zrm}_n=\bmhat{x}$, $\bmhat{\xrm}_n=\bmhat{y}$ and $\bmhat{\yrm}_n=\bmhat{d}$ for all $n$.
 In this case, the coupling is not inversion symmetric $F_{nm}=\left(1+\bmhat{y}\cdot\bmhat{q}\right)^2$, and waves moving in the $\bmhat{y}$ direction couple more strongly than those moving in the $-\bmhat{y}$ direction.
 Finally, if the orientation of the NVs depends on position, the matrix $F_{nm}$ will generally be complex but Hermitian.

\subsubsection{Position dependence of non-local decay rates}
In the case $\bmhat{\zrm}_n=\bmhat{d}$, the integral over the azimuth can be performed and yields
\begin{eqnarray}~\label{eq:decayrates}
    \gamma(r)&=&\left(\frac{\gamma_e}{2}\right)^2\int \frac{qdq}{2\pi}J_0\left(qr\right) C^{-+}\left(\Delta,q\right), 
\end{eqnarray}
where $J_0(x)$ is the zeroth Bessel function of the first kind, and $r=\left|\bm{r}_n-\bm{r}_m\right|$ is the distance between the two dipoles.
The numerical integration of this equation is shown in Fig.~2(b) of the main text. The local decay rate is given by
\begin{align}
    \gamma_0=\gamma(0)\approx\frac{\Delta \gamma_e^2 e^2 k^2_F}{2c^2}\begin{cases}\pi \quad \text{for} \quad d\ll \lambda_F\\
    (k_Fd)^{-1} \quad \text{for} \quad d\gg \lambda_F
    \end{cases}.
\end{align}

\section{Superradiance}~\label{sec:coremiss}
Correlated emission occurs when the nonlocal decay rates $\gamma_{nm}$ with $n\neq m$ are finite.
Depending on the initial state, the relaxation process can be slower (subradiance) or faster (superradiance) than for independent emission.
We will now determine the conditions for superradiance.
To this end, we formulate the adjoint master equation for operators $O$ 
\begin{equation}\label{eq:adjointLindblad}
    \partial_{t}O={\cal L}^\dagger O= \frac{i}{\hbar}{[H_0+\delta H,O]}+ {\cal D}_+ O+{\cal D}_- O+ {\cal D}_\phi O, \end{equation}
where the superoperators are defined by
\begin{align}
 {\cal D}_- O&=\sum_{nm}\gamma_{nm}\left(\sigma^+_n O \sigma^-_m -\frac{1}{2}\{\sigma^+_n \sigma^-_m,O\} \right),\\
  {\cal D}_+O&=\sum_{nm}\bar{\gamma}_{nm}\left(\sigma^-_n O \sigma^+_m -\frac{1}{2}\{\sigma^-_n \sigma^+_m,O\} \right),\\
 {\cal D}_\phi O&=\sum_{nm}\gamma^\phi_{nm}\left(\sigma^z_n O \sigma^z_m -\frac{1}{2}\{\sigma^z_n \sigma^z_m,O\} \right).
\end{align}
Compared to the Letter, we present a more complete master equation describing collective decay with rate $\gamma_{nm}$, collective absorption with rate $\bar{\gamma}_{nm}$, and additional nonlocal dephasing with rate $\gamma^\phi_{nm}$, respectively. The latter can, for example, originate from coupling the spins to classical white noise via $H_\phi=\frac{\hbar}{2}\sum_n \sigma_n^z \eta_n$, where $\ev{\eta_n (t) \eta_m(t')}=4\gamma^\phi_{nm}\delta(t-t')$ and $\ev{\eta_n(t)}=0$. Furthermore, $H_0=\frac{\hbar\Delta}{2}\sum_n\sigma^{\zrm}_n$ is the system Hamiltonian, where we assume that all NV axes are perpendicular to the plane $\bmhat{z}_n=\bmhat{d}$.  Finally, $\delta H$ describes coherent interactions mediated by the environment given by
\begin{align}
\delta H
=
\hbar \sum_{nm}
S_{nm}\,
\sigma_n^{+}\sigma_m^{-} \quad \text{with}\quad S_{nm}
=
\frac{\gamma_e^2}{8\pi}\,
\mathcal{P}
\int_{-\infty}^{\infty}
d\omega
\,
\frac{C^{-+}(\omega,\mathbf{r}_n-\mathbf{r}_m)}
{\Delta-\omega},
\end{align}
where the coefficients $S_{nm}$ follow via a Kramers-Kronig relation. 
Due to the secular approximation, we have $[H_0,\delta H]=0$~\cite{Schaller2014}.

Next, by studying the time dependence of the collective spin $\bm{S}=\frac{\hbar}{2}\sum_n\boldsymbol{\sigma}_n$, we will derive requirements for  superradiant enhancement. We will discuss two separate cases, where the ensemble is prepared in the initial state with $\ev{\bm{S}}=\hbar N \hat{\boldsymbol{z}}/2$ and $\ev{\bm{S}}=\hbar N \hat{\boldsymbol{x}}/2$, respectively.

\subsubsection{Initial state: $\ket{\psi}=\ket{+}^{\otimes N}$}
Assume an initial state $\ket{\psi}=\ket{+}^{\otimes N}$ fully polarized in the excited state, where $\sigma^z\ket{\pm}=\pm\ket{\pm}$. The collective spin is initially pointing in the $z$ direction, $\ev{\bm{S}}=\hbar N \hat{\boldsymbol{z}}/2$. For such an initial state, the initial decay rates for independent and collective emission are the same and given by 
\begin{align}
    R=-\frac{\mel{\psi}{{\cal L}^\dagger S_z}{\psi}}{\hbar}=\sum_n\gamma_{nn}.
\end{align}
Instead, collective emission is distinguished by the initial acceleration of the decay rate as many-body coherences develop.
We find
\begin{align}
    \dot{R}=-\frac{\mel{\psi}{\left({\cal L}^\dagger\right)^2 S_z}{\psi}}{\hbar}=\sum_{n\neq m}\left(1-e^{-\hbar \Delta/kT}\right)\gamma_{nm}\gamma_{mn}-\sum_{n}\left(1+e^{-\hbar \Delta/kT}\right)(\gamma_{nn})^2, \label{eq:dtRgeneral}
\end{align}
where we assumed a thermal environment, such that $\bar{\gamma}_{nm}=e^{-\hbar\Delta/kT}\gamma_{mn}$.  We remark that coherent effects $\delta H$ and dephasing $\gamma^\phi_{nm}$ can and do, in general, interrupt superradiance. 
However, the initial acceleration $\dot{R}$ remains insensitive to these effects. 

For independent emission, the decay-rate matrix $\gamma_{nm}=\gamma_{0}\delta_{nm}$ yields $\dot{R}=-N\gamma_{0}^2(1+e^{-\hbar \Delta/kT})$, and implies the decay process slows as time evolves.
In contrast, correlated emission will occur for a constant decay-rate matrix $\gamma_{nm}=\gamma_{0}$ for all $n$ and $m$. For this case, the decay process accelerates in time as $\dot{R}=N^2\gamma_{0}^2(1-e^{-\hbar \Delta/kT})-2N\gamma_0^2$. For a substantial acceleration, we need $N\gg n$, where $n=\frac{1}{e^{\hbar \Delta/kT}-1}$ is the Bose-Einstein distribution. 
The limit of $\gamma_{nm}=\gamma_{0}$ was originally discussed by Dicke~\cite{PhysRev.93.99}, and leads to a maximum decay rate $\max_t R\propto N^2$ known as the superradiance burst.

\subsubsection{Initial state: $\ket{\psi}=\left(\frac{\ket{+}+\ket{-}}{\sqrt{2}}\right)^{\otimes N}$}
An alternative requirement for collective effects can be observed when the magnetic dipoles are already prepared in a state with coherence between the excited and ground state (used in Ref.~\cite{mokNATPHYS26} to estimate a lower bound on the maximum decay rate). For an initial state $\ket{\psi}=\left(\frac{\ket{+}+\ket{-}}{\sqrt{2}}\right)^{\otimes N}$ with $\ev{\bm{S}}=\hbar N \hat{\boldsymbol{x}}/2$, we obtain 
\begin{align}
    R=-\frac{\mel{\psi}{{\cal L}^\dagger S_z}{\psi}}{\hbar}=\frac{1-e^{-\hbar \Delta/kT}}{2}\left(\sum_k\gamma_{kk}+\frac{1}{2}\sum_{n\neq m}\gamma_{nm}\right).
\end{align}
While independent emission, $\gamma_{nm}=\gamma_{0}\delta_{nm}$, yields ${R}=N\gamma_{0}(1-e^{-\hbar \Delta/kT})/2$, the Dicke limit, $\gamma_{nm}=\gamma_{0}$, leads to ${R}=\gamma_{0}(1-e^{-\hbar \Delta/kT})N(N+1)/4$ scaling like $R\propto N^2$.
Hence, for a state with initial coherence, the initial decay rate $R$ is already sufficient to observe the superradiant enhancement. 
We remark that for a bosonic environment, where $\gamma_{nm}(T)=(n+1)\gamma_{nm}(0)$, the initial decay rate $R$ becomes temperature independent. 

In the following, we consider the zero-temperature limit $T=0$.

\subsection{Comparison to second-order photon-correlation function}
In Ref.~\cite{Masson:2022aa}, the authors use the second-order photon-correlation function $g^2(t=0)$ at the initial time as a condition for superradiance.
They provide the second-order photon-correlation function as
\begin{equation}\label{eq:comparison}
    g^2(t=0)=\frac{\sum_{\nu,\mu}\gamma_\nu\gamma_\mu \left<O_\nu^\dagger O_\mu^\dagger O_\mu O_\nu \right>}{\left(\sum_{\nu}\gamma_\nu \left<O_\nu^\dagger O_\nu \right>\right)^2},
\end{equation}
where $\gamma_\nu$ are the eigenvalues of the decay-rate matrix $\gamma_{nm}$, and $O_{\nu}=\sum_{m}\alpha_{\nu,m}\sigma^-_m$ are the diagonal jump operators determined by the $\nu$th eigenvector $\alpha_{\nu,m}$ of the decay-rate matrix.
Their condition for superradiance, that the emission of the first photon is positively correlated with that of the second $g^2(t=0)>1$, is equivalent to ours $\dot{R}>0$.  This is proven algebraically by substitution of $O_{\nu}=\sum_{m}\alpha_{\nu,m}\sigma^-_m$ into Eq.~\eqref{eq:comparison}, and by using the relation $\gamma_{nm}=\sum_\nu\gamma_\nu\alpha_{\nu,n}\alpha^*_{\nu,m}$, to obtain the expression in Eq.~\eqref{eq:dtRgeneral} appearing on the right-hand side of the inequality of $\dot{R}>0$.  In addition, we used the relation $\left<\sigma^+_{r}\sigma^+_{n}\sigma^-_m\sigma^-_s\right>=\left(\delta_{nm}\delta_{rs}+\delta_{ns}\delta_{rm}\right)\left(1-\delta_{ms}\right)$

\subsection{Homogeneous distribution}
We now consider dipoles distributed homogeneously in a two-dimensional region of size $A$ and at a fixed height $d$.
\subsubsection{Acceleration of decay process $\dot{R}$}
The change in the decay rate $\dot{R}$ depends on the two terms in Eq.~\eqref{eq:dtRgeneral}.
The term $-\sum_{n}\gamma_{nn}^2=-N\gamma_{0}^2$ does not depend on the distribution of dipoles for a homogeneous magnetic-field environment.
The term, $\sum_{n\neq m}\left|\gamma_{nm}\right|^2$ does, and when treating the magnetic-field noise in momentum space, the dependence shows up in a factor similar to the form factor in scattering theory:
\begin{eqnarray}
    \sum_{n\neq m}\left<e^{i(\bm{q}_1-\bm{q_2})\cdot(\bm{r}_n-\bm{r}_m)}\right>=\frac{N^2}{A}\int d^2\bm{r}e^{i\bm{r}\cdot(\bm{q}_1-\bm{q}_2)}g(r),
\end{eqnarray}
where $g(r)$ is the pair-correlation function.
At low density, the NVs are spaced far enough apart that $g(r)\approx 1$.
In this limit, the integral over position can be taken and yields an initial change in the decay rate as
\begin{eqnarray}
    \dot{R}\approx N\left(\frac{\gamma_e}{2}\right)^4\int\frac{d^2\bm{q}_1d^2\bm{q}_2}{(2\pi)^4} C^{-+}(\omega,\bm{q}_1) C^{-+}(\omega,\bm{q}_2)\left(\frac{N}{A}(2\pi)^2\delta(\bm{q}_1-\bm{q}_2)-1\right).
\end{eqnarray}

Upon rearrangement, this can be simply expressed as
\begin{eqnarray}
    \frac{\dot{R}}{N\gamma_{0}^2} = \pi \lambda_{\text{SR}}^2n-1
\end{eqnarray}
where $n=N/A$ is the two-dimensional density of dipoles; we have assumed isotropic noise $C^{-+}(\omega,\bm{q})=C^{-+}(\omega,q)$, and define $\lambda_{\text{SR}}$ as the length scale at which correlated emission occurs:
\begin{eqnarray}
    \pi\lambda_{\text{SR}}^2=\frac{1}{C^{-+}(\Delta)^2}\int \frac{q dq}{2\pi} {C}(\Delta,q)^2,
\end{eqnarray}
where $C^{-+}(\Delta)=\int q dq C^{-+}(\Delta,q)/(2\pi)$ is the local spectral density at frequency $\Delta$.

For the homogeneous distribution, the acceleration $\dot{R}/N\gamma_0^2$ is thus larger by an amount $n\pi\lambda_{\text{SR}}^2$, equal to the number of particles in the correlation region of size $\pi\lambda_{\text{SR}}^2$.
This is in contrast to Dicke superradiance, where the acceleration is increased by the total number of particles $N$.
For concreteness, consider the simple model for magnetic noise $C^{-+}(\Delta,\bm{q}_1) =C\theta(q_{max}-\left|\bm{q}\right|)$ in which magnet noise is cut off above the scale $q_{max}$.
This yields the correlated emission length scale $\lambda_{\text{SR}}^2=4q_{max}^{-2}$, as might be expected.

\subsubsection{Decay rate}
Determining the decay rate for the dipoles initialized in a $\sigma^x$ eigenstate is more subtle.
For a homogeneous environment, we find
\begin{eqnarray}
    R\approx\frac{N}{2}\gamma_{0}+\frac{N^2}{4A}\int d^2\bm{r} g(r)\gamma(r),
\end{eqnarray}
where $g(r)$ is again the pair-correlation function, and the approximation becomes better in the large $N$ limit. For a homogeneous low-density distribution, we approximate $g(r)\approx 1$ in an infinite volume and obtain
\begin{eqnarray}
    R=\frac{N}{2}\gamma_{0}+\frac{N}{4}n\left( \frac{\gamma_e}{2} \right)^2C^{-+}\left( \Delta,\bm{q}=0 \right).
\end{eqnarray}
The second term vanishes if the magnetic-field correlations lack a $\bm{q}=0$ component at finite frequency $\Delta$.
In this case, there is no collective enhancement due to a cancellation from oscillatory contributions from dipoles with large separation.
For realistic configurations, disorder in both NV location and the nearby material will prevent this exact cancellation. 
Therefore, we consider how the rate scales with the area $A$ occupied by the dipoles:
\begin{eqnarray}
    R\approx\frac{N}{2}\gamma_{0}+\frac{N\pi n}{2}\int_0^{\sqrt{A}} rdr \gamma(r).
\end{eqnarray}
As an example, we consider $\gamma(r)=\gamma_0(r_0/r)^{\alpha}$ above a scale $r_0$, and constant below that scale.
For $\alpha<2$, the dominant term grows with $A$ as
\begin{eqnarray}
    R\approx\frac{N\pi n}{2}\frac{A^{1-\alpha/2}r_0^{\alpha}}{2-\alpha}\gamma_0 ,
\end{eqnarray}
and there will be a superradiant burst.
Otherwise, for $\alpha>2$, the total decay rate is similar to the independent rate $R\approx N\gamma_0/2$. 
In two dimensions, transverse current fluctuations generate magnetic-field noise with $\alpha=1$ as discussed in the main text.

\section{Subradiance}

\begin{figure}[b]
    \centering
    \includegraphics[width=4.388in]{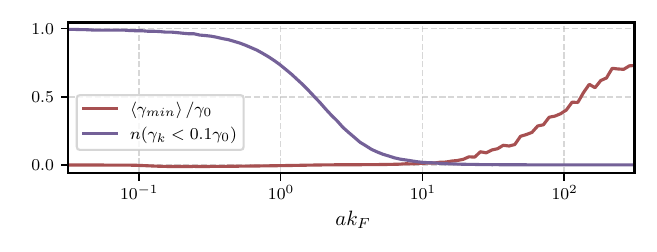}
    \caption{Statistics of single-excitation decay rates $\gamma_k$~(eigenvalues of the decay-rate matrix $\gamma_{nm}$). The figure shows both $\left<\gamma_{min}\right>/\gamma_0$, the smallest decay rate in the system averaged over disorder realizations, and $n(\gamma_k<0.1\gamma_0)$ the proportion of eigenvalues that are smaller than one tenth of the independent decay rate $\gamma_0$. The decay rates were evaluated for $N=500$ dipoles, and the plotted quantities were averaged over $50$ disorder realizations.}
    \label{fig:subradiance_data}
\end{figure}
Subradiance is most well understood in the single-excitation sector.
Here, the density matrix is described by a single-excitation wave function $\ket{\psi(t)}$ and the ground state:
\begin{eqnarray}
    \rho(t)=\ket{\psi(t)}\bra{\psi(t)}+P_0(t)\ket{0}\bra{0}
\end{eqnarray}
where $\ket{0}$ is the $-1$ eigenstate for all $\sigma^z_n$ operators, and the single-excitation wave function has the form
\begin{eqnarray}
    \ket{\psi(t)}=\sum_n \psi_n(t)\sigma^+_n\ket{0},
\end{eqnarray}
with decaying normalization $P_1(t)=1-P_0(t)=\sum_n\left|\psi_n(t)\right|^2$.
Substitution of $\rho(t)$ into the Lindblad equation yields
\begin{eqnarray}
    \partial_t\psi_m(t)= -\frac{1}{2}\sum_{n}\gamma_{nm}\psi_n(t).
\end{eqnarray}

The eigenvalue $\gamma_k$ of the decay-rate matrix $\gamma_{nm}$ is the decay rate of the $k$th single-excitation eigenstate:
The amplitude of the excitation decays as $e^{-\gamma_kt/2}$ such that $P_0(t)=1-e^{-\gamma_k t}$.
The two-dipole decay-rate matrix
\begin{eqnarray}
\gamma_{nm} = \begin{pmatrix}
\gamma_0 & \gamma_a \\
\gamma_a & \gamma_0
\end{pmatrix}
\end{eqnarray}
has eigenvalues $\gamma_0\pm\gamma_a$ corresponding to a dark~($-$) and a bright state~($+$).

\subsection{Homogeneous distribution of spins}

We now consider the case in which a collection of $N$ magnetic dipoles are randomly distributed within a finite square region with density $n=a^{-2}$.
In this case, the single-excitation decay rates are randomly distributed due to disorder in the position of the magnetic dipole locations~(a homogeneous distribution in the square region).
When all magnetic dipoles have a spacing $a\sqrt{N}<\lambda_F$, the decay rate matrix is a constant $\gamma_{nm}\approx \gamma_0$, and the spectrum contains one single-excitation eigenvector with a collective loss rate of $N\gamma_0$ and $N-1$ dark states with loss rate $\approx 0$.

As the spacing increases, some dipoles are separated from each other with separation $r>\lambda_F$, but subradiance remains so long as the average spacing is small $a<\lambda_F$.
Once the average spacing is larger than the Fermi wavelength, subradiance diminishes.
In Fig.~\ref{fig:subradiance_data}, we show subradiance is lost slowly.
For $\lambda_F<a<10\lambda_F$ dark states with $\gamma_k<0.1\gamma_0$ remain in the system, but become more and more rare.
Then for $a>10\lambda_F$, the slowest decaying eigenstate has a small decay but still comparable to the independent rate $\gamma_0$.
Furthermore, for $a>\lambda_F$, a preliminary investigation into the spatial distribution of the dark states shows that the subradiant states are all approximately singlets between two dipoles with separation $r<\lambda_F$.
Since the mean spacing $a>\lambda_F$ is large, these states are rare but still occur at a finite density decreasing with increasing mean spacing $a$.

\section{Magnetic moment}
In this section, we review the physics of correlated emission in the Dicke limit, where the decay-rate matrix is constant,  $\gamma_{nm}=\gamma_0$.
The dynamics of the collective spin operator $\bm{S}=\frac{\hbar}{2}\sum_n \boldsymbol{\sigma}_n$ is obtained from the adjoint Lindblad equation from Eq.~\eqref{eq:adjointLindblad}. Defining $S^\pm=S^x\pm i S^y$, we obtain the coupled equations in the rotating frame:
\begin{equation}
\begin{split}
    \hbar\partial_t S^z &= -\gamma_0 S^+ S^-,    \\
    \hbar\partial_t S^+ &=  \gamma_0 S^+ S^z,    \\
    \hbar\partial_t S^- &=  \gamma_0 S^z S^-.
\end{split}
\label{eqn: collective dynamics}
\end{equation}
This equation can be rewritten into the vector form:
\begin{align}
\hbar\partial_t \bm{S} 
= -\gamma_0 \bm{S}\times \left(\hat{\bm{z}} \times \bm{S} \right) - \hbar\frac{\gamma_0}{2} \left(\hat{\bm{x}}- i \hat{\bm{y}}\right) S^+ -\hbar\gamma_0 \hat{\bm{z}}S^z,
\end{align} 
\begin{align}
\hbar\partial_t \bm{S} 
= - \gamma_0\left[ \bm{S}\times \left(\hat{\bm{z}} \times \bm{S} \right)\right]_{\text{sym}} - \hbar\frac{\gamma_0}{2} \left(\bm{S}+ \hat{\bm{z}}S^z\right) ,
\end{align} 
where the first term can be identified as the symmetrized Landau-Lifshitz damping $\left[ \bm{S}\times \left(\hat{\bm{z}} \times \bm{S} \right)\right]_{\text{sym}}=\left[\bm{S}\times \left(\hat{\bm{z}} \times \bm{S} \right)- \left(\hat{\bm{z}} \times \bm{S} \right)\times \bm{S}\right]/2$, which becomes dominant for large spins, $N\gg1$. Transforming back to the lab frame and using the mean-field approximation, we obtain for the expectation value
\begin{align}
\partial_t \ev{\bm{S} }
\approx \Delta \hat{\bm{z}} \times \ev{\bm{S}}  -\frac{\gamma_0}{\hbar} \ev{\bm{S}}\times \left(\hat{\bm{z}} \times \ev{\bm{S}} \right).
\end{align}
Here, we used that for large spins, $N\gg1$, Landau-Lifshitz damping $\propto N^2$ dominates. The exact solution to this equation is given by
\begin{equation}
\begin{split}
    \ev{S^x}&=\frac{N\hbar}{2}\cos(\Delta t) \sech(\frac{N\gamma_0}{2} t),\\
        \ev{S^y}&=\frac{N\hbar}{2}\sin(\Delta t) \sech(\frac{N\gamma_0}{2} t),\\
            \ev{S^z}&=- \frac{N\hbar}{2}\tanh(\frac{N\gamma_0}{2} t),
\end{split}\label{eq:solutionLandauLifshitz}
\end{equation}
where the initial condition is fully polarized in the $\hat{\bm{x}}$ direction, $\ev{\bm{S}(t=0)}=\left(\frac{N\hbar}{2},0,0\right)$.

\section{Current response}
In this section, we derive the electron current response $\bm{J}(\bm{r},t)$ to the magnetic field generated by the ensemble of magnetic dipoles.  A movie illustrating the time evolution of the current as the spin ensemble decays is provided at \cite{SM2}.
As above, we only consider the transverse current $J_\perp(\bm{q},\omega)=\hat{\bm{q}}_\perp\cdot\bm{J}(\bm{q},\omega)$, which we determine via the Kubo formula~\cite{Coleman15} for the transverse-current response $J_\perp$ to the magnetic field $\nabla\times\bm{A}$ : 
\begin{align}
\ev{J_\perp(\bm{q},\omega)}=-\left[\frac{1}{\hbar}G^R_0(\bm{q},\omega)+\frac{e^2\rho_e}{m}\right]\hat{\bm{q}}_\perp\cdot\bm{A}(\bm{q},\omega)/c,\label{eq:Kubo}
\end{align}
where the first term is the ``paramagnetic'' current $\bm{J}^\text{para}$ given by the retarded Green's function $G^R_0$ of current-current fluctuations and the second term is the ``diamagnetic'' current $\bm{J}^\text{dia}$,
and $\rho_e=k_F^2/2\pi$ is the 2D electron density.

The vector potential $\bm{A}$ in the plane of the 2D electron gas can be derived from the magnetic interaction $H_\text{int}=-\bm{m}\cdot\bm{B}=-\int{\frac{d^2\bm{q}}{(2\pi)^2}\,\bm{J}(-\bm{q})\cdot\bm{A}}(\bm{q})/c$. Using Eq.~\eqref{eq:magpropcross}, we obtain
\begin{equation}
  \bm{A}(\bm{q},\omega) = - c\bm{G}(-\bm{q})\times\bm{m}(\omega),
\end{equation}
where $\bm{G}$ is the magnetostatic Green's function defined in Eq.~\eqref{eqn: Biot-Savart}, $\bm{m}(\omega)= -\gamma_e \expval{\bm{S}(\omega)}$ is the magnetic moment of the decaying macrospin, and we have applied the Coulomb gauge $\nabla\cdot\bm{A}=0$. 
The evolution of the macrospin $\expval{\bm{S}(\omega)}$ is given by the Fourier transform of Eq.~\eqref{eq:solutionLandauLifshitz}, where we assume the spin remains polarized along the positive $\hat{\bm{x}}$ direction for $t<0$ and subsequently evolves for $t\geq 0$.

In total the response of the current to the decaying macrospin is written as 
\begin{align}
\expval{J_\alpha(\bm{r},t)} = \int \frac{d\omega}{2\pi}e^{-i\omega t}  \chi_{\alpha\beta}(\bm{r},\omega)\ev{S_\beta(\omega)},
\label{eqn: current response}
\end{align}
where $\alpha\in\{x,y\}$ describes the two-dimensional conductor and $\beta\in\{x,y,z\}$ the spin direction.
The response function $\chi_{\alpha\beta}$ is found as:
\begin{equation}
\begin{split}
    \chi_{\alpha\beta}(\bm{r},\omega) 
    &=\frac{2\pi \gamma_e}{ c} \int{\frac{d^2\bm{q}}{(2\pi)^2}e^{i\bm{q}\cdot\bm{r}} (\hat{\bm{q}}_{\perp})_\alpha(-\hat{\bm{q}}+i\hat{\bm{d}})_\beta\, e^{-q d}\left[\frac{1}{\hbar}G^R_0( \bm{q},\omega)+\frac{e^2\rho_e}{m}\right]},
    \label{eqn: chi1}
\end{split}
\end{equation}
where we assumed $d>0$ and used the right-handed basis $\hat{\bm{q}}_{\perp},\hat{\bm{q}}, \hat{\bm{d}}$.
Next, we assume an isotropic response $G_0^{R}(\bm{q},\omega)=G_0^{R}(q,\omega)$, and perform the integration over the azimuthal angle of $\bm{q}$.
We arrive at
\begin{equation}
\begin{split}
    \chi_{\alpha\beta}(\rho,\varphi,\omega) =\frac{2\pi \gamma_e}{ c}  \int_0^{+\infty}{\frac{qdq}{2\pi}K_{\alpha\beta}(q\rho,\varphi)e^{-q d}\left[\frac{1}{\hbar}G^R_0(q,\omega)+\frac{e^2\rho_e}{m}\right]}, 
    \label{eqn: chi2}
\end{split}
\end{equation}
where we parametrize $\bm{r}$ by polar coordinates $\rho$ and $\varphi$. Here, we define  
\begin{equation}
  K_{\alpha\beta}(q\rho,\varphi)=
  \begin{pmatrix}
    A(\rho q)\sin{\varphi} & -A(\rho q)\cos{\varphi}  & 0 \\
    B(\rho q)\cos{\varphi} & B(\rho q)\sin{\varphi} & J_1(q\rho)
  \end{pmatrix}_{\alpha\beta},
  \label{eqn: Bessel functions}
\end{equation}
where $A(\rho q) = \frac{1}{2}\left[J_0(q\rho)+J_2(q\rho)\right]$, $B(\rho, q) = \frac{1}{2}\left[J_0(q\rho)-J_2(q\rho)\right]$ and $J_n$ are Bessel functions of the first kind.

\subsection{Current-current response}
We calculate the current-current response $G_0^R$ for the free Fermi gas with dispersion relation $\omega_{\bm{k}}=\hbar \bm{k}^2/2 m$ and at zero temperature:
\begin{equation}
\begin{split}
    G^R_0(\bm{q}, t) &=-i \Theta(t)\frac{1}{{\cal A}}\ev{\left[\hat{\bm{q}}_\perp\cdot\bm{J}(\bm{q},t),\hat{\bm{q}}_\perp\cdot\bm{J}(-\bm{q},t) \right]}_0 \\
    &= -i\Theta(t)\frac{1}{{\cal A}}\left(\frac{e \hbar }{m}\right)^2\sum_{\sigma,\bm{k}}\left(\hat{\bm{q}}_{\perp}\cdot\bm{k}\right)^2\expval{[f^{\dagger}_{\sigma,\bm{k}}(t)f_{\sigma,\bm{k}+\bm{q}}(t), f^{\dagger}_{\sigma,\bm{k}+\bm{q}}f_{\sigma,\bm{k}}]}_0,\label{eq:retardedG}
\end{split}
\end{equation}
where ${\cal A}$ is the area of the 2D plane, and in the second step we inserted Eq.~\eqref{eqn: current operator}.
The current response $G^R_0(\bm{q}, t)$ and the corresponding Fourier components $G^R_0(\bm{q}, \omega)$ can be calculated exactly in two dimensions: 
\begin{align}
    G^R_0(q,t) &= -\frac{4\hbar e^2\rho_e}{m}k_F v_F \Theta(t)\sin\left(\frac{q}{2k_F}q v_F t\right)\frac{J_2(q v_F t)}{(q v_F t)^2},  \\
    G^R_0(q,{\omega}) &= \frac{\hbar e^2\rho_e}{m}\frac{2 k_F}{3q}\left[g\left(\frac{\omega}{v_Fq}+\frac{q}{2k_F}\right)-g\left(\frac{\omega}{v_F q}-\frac{q}{2 k_F}\right)\right], \label{eq:retardedGFsol} 
\end{align}
where $g(x)$ is a complex piecewise function:
\begin{equation}
\begin{split}
    \Re[g(x)] &\equiv x^3-\frac{3}{2}x - (x^2-1)^{\frac{3}{2}}\Theta(x^2-1)\text{sign}(x)   \\
    \Im[g(x)] &\equiv (1-x^2)^{\frac{3}{2}}\Theta(1-x^2).
\end{split}
\end{equation}
It can be shown that the real and imaginary parts of $g(x)$ satisfy the Kramers-Kronig relations.
The Green's function simplifies under the long-wavelength approximation $q\ll k_F$ to 
\begin{align}
    G^R_0({{q}}, {\omega}) 
    &\approx  \frac{e^2\rho_e}{m}\frac{2\hbar}{3}g^\prime \left(\frac{\omega}{v_Fq}\right), \label{eqn: retarded approx}
\end{align}
where $g^\prime(x)=\partial_x g(x)$. 

We remark that the imaginary part of the current response $\Im G_0^R$ from Eq.~\eqref{eq:retardedGFsol}  (capturing dissipation) and the magnetic noise $C^{-+}$ from Eq.~\eqref{eq:noisepm} (originating from current fluctuations) are related to each other via the fluctuation-dissipation theorem at zero temperature for $\omega>0$:
\begin{align}
    C^{-+}(\omega,q)=-2\,\mathrm{Im}G^R_0(q,\omega)\left(\frac{2\pi}{c} e^{-qd}\right)^2,
\end{align}
where we used Eq.~\eqref{eqn: Biot-Savart}.

\subsection{Dynamic response}
The current response $\bm{J}(\bm{r},t)$ can be obtained by numerical integration of Eq.~\eqref{eqn: current response} and Eq.~\eqref{eqn: chi2}. 
The resulting current distribution is shown in Fig.~3 of the main text.
We can see that $\bm{J}$ travels away from the macrospin $\ev{\bm{S}}$ with velocity $v_F$, while the amplitude decays with distance. 
The current distribution forms the pattern of Archimedean spirals with two arms merging at the origin (the position of the macrospin) in a way that ensures  $\grad \cdot \bm{J} =0$.
In addition, loop currents are superimposed on these spirals, which favor an overall clockwise current direction (in accordance with Lenz's law).

To better understand these features of the current $\bm{J}=\bm{J}^\text{para}+\bm{J}^\text{dia}$, we perform the far-field approximation using the following length scale hierarchy~\footnote{In principle, there is also an upper length scale $\rho\ll \frac{E_F}{\hbar \abs{\omega}}\frac{v_F}{\abs{\omega}}$ used in our approximation, which stems from the splitting of two nearby particle-hole thresholds neglected in the long-wavelength approximation.
However, for typical values, this length scale is expected to be much larger than the sample size.}
\begin{align}
    \text{max}(d,\lambda_F)\ll \frac{v_F}{\abs{\omega}}\ll \rho.
\end{align}
In this limit, the paramagnetic current $\bm{J}^\text{para}$ is described by waves propagating at the Fermi velocity $v_F$ arising from the low-momentum edge of the particle-hole continuum. The diamagnetic current $\bm{J}^\text{dia}$ is given by the instantaneous response to the vector potential $\bm{A}$.
In the end, we can show that the response is given by:
\begin{align}
     \chi^\text{para}_{\alpha\beta}(\rho,\varphi,\omega) 
     &\approx \frac{2e^2\rho_e\gamma_e}{m c} 
     \frac{e^{i\frac{\rho\omega}{v_F}}}{\rho^2}
    \begin{pmatrix}
        -\frac{iv_F}{\omega \rho}\sin{\varphi} & \frac{iv_F}{\omega \rho}\cos{\varphi} & 0   \\
        \cos{\varphi} &  \sin{\varphi} & 
        -i\,\text{sign}(\omega)
     \end{pmatrix}_{\alpha\beta},    \\
     \chi^\text{dia}_{\alpha\beta}(\rho,\varphi)
     &\approx -\frac{e^2\rho_e\gamma_e}{m c} \frac{1}{\rho^2}
     \begin{pmatrix}
         -\sin{\varphi} & \cos{\varphi} & 0  \\
         \cos{\varphi} & \sin{\varphi} & -1
     \end{pmatrix}_{\alpha\beta}.
\end{align}
The validity of these approximations has been checked by numerical integration. Note that the paramagnetic response function has a phase factor $e^{i\rho\,\omega/v_F}$, which gives rise to the traveling-wave form, so that $\bm{J}^\text{para}$ depends on the delayed time $t-\rho/v_F$. Thus, $\bm{J}^\text{para}$ captures ballistic current waves.

Inserting the response function $\chi^\text{para}_{\alpha\beta}$ into Eq.~\eqref{eqn: current response} and using Eq.~\eqref{eq:solutionLandauLifshitz}, 
we find for the oscillatory part of the paramagnetic current  $\bm{J}^\text{para}$, which is driven by $\ev{S^x}$ and  $\ev{S^y}$, the following result 
\begin{equation}
\begin{split}
    J^\text{para}_{\rho}(\rho,\varphi,t) &\approx -K \frac{v_F}{\Delta\rho^3}\Theta\left(t-\frac{\rho}{v_F}\right)\sech\left[\frac{N\gamma_0}{2} \left(t-\frac{\rho}{v_F}\right)\right]\cos\left[\Delta\left(t-\frac{\rho}{v_F}\right)-\varphi\right],   \\
    J^\text{para}_{\varphi}(\rho,\varphi,t) &\approx K \frac{1}{\rho^2}\Theta\left(t-\frac{\rho}{v_F}\right)\sech\left[\frac{N\gamma_0}{2} \left(t-\frac{\rho}{v_F}\right)\right]\cos\left[\Delta\left(t-\frac{\rho}{v_F}\right)-\varphi\right],
\end{split}
\end{equation}
where the prefactor is given by $K=N\hbar e^2\rho_e\gamma_e/mc$. Here, we assumed  $N\gamma_0\ll \Delta$, which is consistent with the Markov approximation.  At long distances, the radial component scales as $J^\text{para}_\rho\propto\rho^{-3}$ and the azimuthal component scales as $J^\text{para}_{\varphi}\propto\rho^{-2}$. The Archimedean spirals are visible from the constant-phase curve $\Delta(t-\rho/v_F)-\varphi=\text{constant}$, producing an outward propagating alternating current with radial spatial period $2\pi v_F/\Delta$.

In addition to the spirals that stem from the oscillatory source terms $\ev{S_x}$ and $\ev{S_y}$, there are loop currents superimposed on the spirals induced by the nonoscillatory source term $\ev{S_z}$. 
Furthermore, the total current shown in Fig.~3 also contains the diamagnetic contribution $\bm{J}^\text{dia}$ given in Eq.~\eqref{eq:Kubo}, whose dynamics are directly determined by the vector potential.

\subsection{Static response}
Finally, we show that the static response agrees with the well-known result of Landau diamagnetism.
By relating the loop currents to a magnetization, ${m}^d_J= \frac{i}{cq} \hat{\bm{d}}\cdot \left( \hat{\bm{q}}\times \ev{\bm{J}}\right)$, and the vector potential to the magnetic field, $H^d= i \hat{\bm{d}}\cdot \left({\bm{q}}\times \bm{A}\right)$,  we find from Eq.~\eqref{eq:Kubo} the magnetic response, ${m}^d_J =\chi_\text{Landau} {H}^d$.
Hence, we determine the magnetic susceptibility as~\cite{PRBSingh75}
\begin{equation}
    \chi_\text{Landau}=\lim_{\omega\rightarrow 0}\frac{-1}{c^2 q^2}\left(\frac{1}{\hbar}G^R_0(\bm{q},\omega)+\frac{e^2\rho_e}{m}\right)
    =-\frac{e^2}{12\pi m c^2}=-\frac{1}{3}\chi_\text{Pauli} \frac{m_e^2}{m^2},
\end{equation}
where we used Eq.~\eqref{eq:retardedGFsol}. Here,  $m_e$ is the free electron mass, and $m$ the effective band mass of the electron.
In addition, we remark that the spatial dependence of the static response scales as $\ev{\bm{J}}\propto \rho^{-4}$, which follows from Eq.~\eqref{eqn: chi2} for $\omega\rightarrow 0$ and $d\rightarrow 0$. Hence, it decays much quicker than the dynamic response and is not resolved in Fig. 3 of the Letter.

\bibliography{refs.bib}